\documentclass[twocolumn,tighten]{aastex63}

\DeclareFontFamily{OT1}{pzc}{}
\DeclareFontShape{OT1}{pzc}{m}{it}{<-> s * [1.300] pzcmi7t}{}
\DeclareMathAlphabet{\mathpzc}{OT1}{pzc}{m}{it}

\usepackage{amssymb}
\usepackage{amsmath}
\usepackage{mathtools}
\usepackage{graphicx}
\usepackage{natbib}
\usepackage{url}
\usepackage{paralist}
\usepackage{epsfig}
\usepackage{bbm}
\usepackage{hyperref}

\newcommand{\alf}{Alfv$\acute{\text{e}}$n } 
\newcommand{\alfic}{Alfv$\acute{\text{e}}$nic } 
 
\newcommand{\nCRo}{$n_{\mathrm{CR,0}}$ } 

\defcitealias{Bai2019}{BOPS19}

\received{XXXX}
\revised{XXXX}
\accepted{XXXX}

\submitjournal{ApJ}

\shorttitle{MHD-PIC Simulations of Molecular Clouds}
\shortauthors{C.~J.~Bambic, X.~Bai, \& E.~C.~Ostriker}

\graphicspath{{./}{Figures/}}

\begin{document}

\title{Cosmic Ray Transport in the Ionized and Neutral ISM: MHD-PIC Simulations and \\ Effective Fluid Treatments}

\author[0000-0001-5320-9578]{Christopher~J.~Bambic}
\affiliation{Department of Astrophysical Sciences, Peyton Hall, Princeton University, Princeton, NJ 08544, USA; \href{cbambic@princeton.edu}{cbambic@princeton.edu}}

\author[0000-0001-6906-9549]{Xue-Ning~Bai}
\affiliation{Institute for Advanced Study, Tsinghua University, Beijing 100084, People’s Republic of China; \href{xbai@tsinghua.edu.cn}{xbai@tsinghua.edu.cn}}
\affiliation{Department of Astronomy, Tsinghua University, Beijing 100084, People’s Republic of China}

\author[0000-0002-0509-9113]{Eve~C.~Ostriker}
\affiliation{Department of Astrophysical Sciences, Peyton Hall, Princeton University, Princeton, NJ 08544, USA; \href{eco@astro.princeton.edu}{eco@astro.princeton.edu}}

\begin{abstract}
    
Cosmic rays (CRs) have critical impacts in the multiphase interstellar medium (ISM), driving dynamical motions in low-density plasma and modifying the ionization state, temperature, and chemical composition of higher-density atomic and molecular gas. We present a study of CR propagation between the ionized ISM and a neutral cloud. Using one-dimensional magnetohydrodynamic particle-in-cell simulations which include ion-neutral drag to damp \alf waves in the cloud, we self-consistently evolve the kinetic physics of CRs and fluid dynamics of the multiphase gas. By introducing the cloud in our periodic domain, our simulations break translational symmetry and allow the emergence of spatial structure in the CR distribution function. A negative spatial gradient forms across the fully-ionized ISM region while a positive gradient forms across the neutral cloud. We connect our results with CR hydrodynamics formulations by computing the wave-particle scattering rates as predicted by quasilinear, fluid, and Fokker-Planck theory. For momenta where the mean free path is short relative to the box size, we find excellent agreement among all scattering rates. By exploring different cloud sizes and ion-neutral collision rates, we show that our results are robust. Our work provides a first-principles verification of CR hydrodynamics when particles stream down their pressure gradient, and opens a pathway toward comprehensive calibrations of transport coefficients from self-generated \alf wave scattering with CRs.

\end{abstract}


\section{Introduction} \label{sec:intro}

The gas in the interstellar medium (ISM) spans a huge range of temperatures and densities, with the coldest atomic and molecular phases taking the form of clouds embedded within a diffuse medium \citep{Dickey1990,Wolfire1995,Draine2011}. Pervading both the diffuse ISM and dense clouds are relativistic cosmic rays (CRs), with a (broken) power-law distribution extending over more than ten orders of magnitude in energy \citep{Grenier2015}.  With the total energy density in CRs comparable to the thermal, kinetic, and magnetic energy densities of the thermal ISM plasma, CRs may have important dynamical consequences, especially in extraplanar regions where they may help drive galactic winds \citep{Zweibel2017,Recchia2021}.  While CRs of energy $\sim {\rm GeV}$ dominate dynamical effects, the lower-energy portion of the distribution is the most important source of ionization in regions shielded from UV radiation by dust \citep{Padovani2020}. This ionization drives both chemistry and heating. 

Cold atomic and molecular clouds have low ionization fraction ($x_i \lesssim 10^{-4}$; \citealt{Draine2011}), and short-wavelength \alf waves propagating in these clouds are subject to strong damping through ion-neutral collisions. Because CR transport is governed by scattering off of these waves (see below), CR mean free paths become quite long within clouds. Furthermore, since the envelopes of clouds represent transition regions to the more ionized warm, diffuse gas, the propagation of CRs through clouds may depend strongly on the structure of this boundary layer \citep{Padovani2009,Everett2011,Ivlev2018,Silsbee2019}. 

\alf waves in the ISM are subject to growth via the gyroresonant cosmic ray streaming instability \citep[CRSI;][]{Lerche1967,Kulsrud1969}. If the CR drift velocity $v_D$ 
(the mean velocity of the CR distribution in the rest frame of the thermal gas)  exceeds the \alf velocity $v_A$, \alf waves grow rapidly, feeding off the free energy from the momentum space anisotropy of the drifting distribution. Magnetic perturbations exert Lorentz forces on particles, pitch angle scattering those in gyroresonance with \alf waves. The resulting diffusion in momentum space tends to isotropize the CR distribution in the frame of the waves. The CRSI saturates once the momentum distribution function becomes isotropic in the wave frame, implying a drift velocity $v_D \sim v_A$. 

A nearly isotropic CR distribution lends itself to a fluid description. While charged particles generically obey the 6D Vlasov-Maxwell system of equations, self-generated \alf waves limit the anisotropy of the CR distribution function, ensuring that the first 
two moments (energy density and flux)
describe the evolution. In essence, the near isotropy of the drifting distribution provides a closure for a hierarchy of moment equations, transforming the CR pressure tensor into a scalar and making kinetic particles behave as a continuous fluid. 

The fluid approach to CR transport, CR hydrodynamics \citep{McKenzie1982,Zweibel2017}, is particularly useful on astrophysically macroscopic scales. CRSI-generated waves grow on the CR gyroscale ($\sim$10$^{-8}-10^{-6}$~pc) while the overall distribution function varies on much larger scales, $\sim$ 1-10$^4$ pc, a separation of up to 12 orders of magnitude in length scale. CR hydrodynamics overcomes this scale separation by wrapping the microphysics of \alf wave-particle interactions into momentum-dependent ``diffusion coefficients'' which are themselves functions of the wave-particle scattering rate \citep{Shalchi2009,Pfrommer2017}. Unfortunately, computing the diffusion coefficients is nontrivial, and has remained a subject of intense debate with relevance for the solar wind \citep{Jokipii1971}, molecular clouds \citep{Padovani2020}, galaxy formation/evolution  \citep{Buck2020,Hopkins2021a,Hopkins2021b}, galactic winds and outflows \citep{Farber2018,Zweibel2017,Bustard2020}, galactic halos \citep{Kempski2020,Ji2020}, and even the intracluster medium \citep[ICM;][]{Guo2008,Ruszkowski2017}.

Low-ionization regions further complicate transport. In weakly-ionized atomic and molecular clouds, collisions between ions and neutral particles 
impose a drag force on the ions, damping the short-wavelength \alf waves that are gyroresonant with CRs at GeV and lower energies \citep{Kulsrud1969}. Without significant wave-particle scattering, CR transport is ballistic, with particles freely streaming through clouds. In this situation, the CR mean free path to wave-particle scattering can approach or exceed the scale of clouds \citep[][]{Abdo2010}. Thus, the cloud-ISM interface represents a transition in the physics of CR transport.

The non-trivial large-scale CR dynamics induced by this interface has been the subject of a number of previous works. \cite{Skilling1976} proposed that scattering of CRs by their self-generated \alf waves could divert their trajectories, ``excluding'' them from entering molecular clouds with the large fluxes inferred from the diffusive medium. This conclusion is countered by \cite{Cesarsky1978} who argued that diffusion due to small-scale magnetic irregularities and self-generated waves is ineffective at excluding CRs from GMCs, except for particles with energies $\lesssim 50$ MeV. 

\cite{Everett2011} revisited this system using fluid theory with an imposed spatial diffusion coefficient. Using the steady state CR hydrodynamics equations, they found that variation of the diffusion coefficient across the interface could decrease the number density within clouds by an up to an order of magnitude. \cite{Ivlev2018} argued that such modulation can be explained as a ratio of the CR fluxes due to diffusion and free streaming. \cite{Silsbee2019} provided a means of connecting these numerical and semi-analytic models to observations of the CR ionization rate in GMCs \citep{Indriolo2007}, finding that variations in the transport parameters yielded significant changes to the ionization rate vs. column density relation. Recently, \cite{Fujita2020} argued that X-ray and gamma ray observations could be used to constrain the transport regime of CRs in GMCs. In galaxy-scale fluid simulations, \cite{Semenov2020} found that changing CR diffusivity around GMCs acts to regulate star formation and the structure of galactic disks. Simple analytic and fluid models consistently point to observational consequences imposed by variations in CR transport physics.

Fluid models with a constant spatial diffusion coefficient are often used to examine the role of CRs on galaxy properties \citep{Booth2013,Salem2014,Pakmor2016,Pfrommer2017,Wiener2017,Chan2019,Hopkins2021b}. While these models can yield useful constraints and comparisons to observations \citep{Hopkins2021a}, a full treatment of CRs around GMCs requires calculating the diffusion coefficient from first principles.

In this paper, we take the first steps toward a kinetic study of CR propagation through the multiphase ISM. Using magnetohydrodynamic particle-in-cell (MHD-PIC) simulations developed within the {\it Athena} code framework \citep{Bai2015}, we study a drifting population of CRs within a toy model of a magnetized ISM-cloud system: a one-dimensional box with distinct cloud and ISM regions separated by a sharp boundary. Our simulations self-consistently evolve the gyroresonant CRSI and subsequent CR-wave interactions in both regions while including the effects of ion-neutral damping of \alf waves in the cloud. Thus, we extend the simulations of \cite{Bai2019} (hereafter \citetalias{Bai2019}), which studied the gyroresonant CRSI in a uniform plasma without damping, to a model system relevant for the multiphase ISM. In separate work, we have also used MHD-PIC to investigate the effect of ion-neutral drag on development of the CRSI in linear, post-linear, and saturated stages, considering a uniform medium but different levels of the drag coefficient (Plotnikov et al 2021, in prep.).  

With our MHD-PIC simulations, we are able to directly follow CR transport that results from wave-particle scattering, on a particle-by-particle level. Ion-neutral damping in the cloud region results in spatial variation in the \alf wave energy density and thus CR transport properties, with free streaming in the cloud and diffusive propagation in the ambient ISM. By breaking translational symmetry in the simulations, damping in the cloud allows for the emergence of a spatial gradient in the CR number and energy densities. In the ambient ISM, the energy flux is in the direction of the negative energy density gradient, confirming that CRs stream down density gradients \citep{Wentzel1971}.  
We show that the spatial structure in the cloud region (where scattering is negligible) is also consistent with CR hydrodynamics, provided time-dependent terms are included.  

The \alf wave-particle scattering rate encodes the transport properties of the system, and for quantitative comparisons we compute the effective wave-particle scattering rate using three different methods: quasilinear theory based on the \alf wave spectrum, Fokker-Planck theory based on individual particle orbits, and fluid theory based on a moment equation for the CR energy flux. We show that all of these methods yield equivalent scattering rates when the mean free path is sufficiently short, thus providing a first principles verification of CR hydrodynamics in the diffusive regime. Using the resulting \alf wave spectrum, we compute a spatial diffusion coefficient. In this way, our work creates a pathway toward first-principles calculations of the CR spatial diffusion or scattering coefficients, with relevance to ISM structure and thermal/chemical properties,  star formation, galaxy formation, launching of galactic winds, and heating in galaxy halos and the ICM.

This paper is organized as follows. In Section~\ref{sec:theory}, we introduce key theoretical results and different definitions of the wave-particle scattering rate. Section~\ref{sec:methods} details the MHD-PIC method and the numerical set-up for our simulations. Sections~\ref{sec:spatial_structure} and~\ref{sec:scattering_rates} present our main results: the emergence and origin of a spatial gradient in energy density, and a comparison among different calculations of the wave-particle scattering rate for each simulation. In Section~\ref{sec:discussion}, we discuss the validity of the approximations used to compute the wave-particle scattering rates, sources of discrepancy among the rates, limitations in our model, and implications for CR transport in the multiphase ISM. We conclude in Section~\ref{sec:conclusion}.

\section{Theoretical Preliminaries} \label{sec:theory}

The fluid-like behavior of cosmic ray (CR) particles as described by CR hydrodynamics is a consequence of wave-particle scattering from \alf waves excited by the gyroresonant CRSI. In this section, we discuss key results of the CRSI and how CR hydrodynamics is constructed from quasilinear theory. The resulting moment equations will motivate the results in this work, namely the emergence of a spatial gradient in the CR energy density as well as the computation of the wave-particle scattering rate.

\subsection{Wave-Particle Scattering Rates}

Throughout this paper, we work in spherical coordinates in momentum space, $(p,\theta,\phi)$, where $p$ is the CR momentum, $\theta$ is the pitch angle, and $\phi$ is the gyrophase. We define the pitch angle of a particle relative to the magnetic field direction through its cosine, $\mu~\equiv~\cos(\theta)$. With this definition, particles moving forward, parallel to the magnetic field correspond to $\mu = 1$, particles moving backward, parallel to the field have $\mu = -1$, and particles gyrating about the magnetic field completely perpendicular to the field direction have $\mu=0$. This latter case corresponds to a 90$^{\circ}$ pitch angle.

A simple analysis of a particle propagating along a background magnetic field $\boldsymbol{B}_0 = B_0 \hat{x}$ perturbed by an \alf wave with frequency $\omega$ shows that the particle will be in resonance with the wave when the particle momentum $\boldsymbol{p}$ = $\gamma m \boldsymbol{v}$  satisfies
\begin{equation}
    \omega - k v \mu \pm \Omega = 0.
\end{equation}
Here, $k$ represents the wavenumber along the background field, and the $\pm$ symbol represents the difference in resonance condition for left ($+$) and right ($-$) handed waves.
The magnitude of the particle velocity $\boldsymbol{v}$ is denoted by $v$, and $\Omega$ is the relativistic CR gyrofrequency, related to the non-relativistic cyclotron frequency $\Omega_c$ by
\begin{equation}
    \Omega = \frac{\Omega_c}{\gamma} = \frac{q B_0}{\gamma m c},
\end{equation}
where $q$ and $m$ are the CR proton charge and ion mass respectively, and the Lorentz factor $\gamma$ is given by,
\begin{equation} \label{eq:gamma}
   \gamma = \frac{1}{\sqrt{1-\left(
   \frac{|\boldsymbol{v}|}{c} \right)^2}} = 
   \sqrt{1 + \left( \frac{|\boldsymbol{p}|}{mc} \right)^2}.
\end{equation}
Where appropriate, we differentiate between momentum or wave components parallel ($\parallel$) and perpendicular ($\bot$) to the equilibrium field. 
For an \alfic perturbation $\delta \boldsymbol{B}_{\bot}$ such that the amplitude satisfies $|\delta \boldsymbol{B}_{\bot}|$~$\ll$~$|\boldsymbol{B}_0|$, a particle in gyroresonance with the wave will experience an average parallel Lorentz force, leading to a change in pitch angle,
\begin{equation}
    \Delta \theta \approx \pi  \left(\frac{|\delta \boldsymbol{B}_{\bot}|}{|\boldsymbol{B}|} \right) \cos{\phi},
\end{equation}  
over one cyclotron orbit \citep{Kulsrud2005}. Within the quasilinear approximation, uncorrelated \alf wave packets each scatter a resonant particle by $\Delta \theta$, leading to random walk diffusion of the pitch angle at a rate,
\begin{equation} \label{eq:naive_scattering}
    \nu_{\mathrm{scat}} = \sum_{i=1}^{N} 
    \frac{\langle \Delta \theta^2 \rangle_i}{2 \Delta t} = \frac{N (\Delta \theta)^2}{8\pi N\Omega^{-1}} = \frac{\pi}{8} 
    \left(\frac{\delta B_{\bot}}{B} \right)^2 \Omega,
\end{equation}
where we sum over the number of scattering events $N$.

Working out the full quasilinear theory refines this scattering rate. On slow timescales, a particle distribution $f(t,x,\boldsymbol{p})$ in a background magnetic field under the influence of a spectrum of weak electromagnetic fluctuations undergoes quasilinear diffusion \citep{Kennel1966}. In the frame comoving with \alf waves, the wave electric field vanishes since the magnetic field is stationary in this frame. Because magnetic fields can do no work on the particles, diffusion in momentum space must occur through pitch angle scattering \citep{Jokipii1966}. Thus, particles obey the quasilinear diffusion equation,
\begin{equation} \label{eq:quasilinear_diffusion}
   \frac{\partial f_w}{\partial t} + 
    v_w \mu_w \frac{\partial f_w}{\partial x} = 
    \frac{\partial}{\partial \mu_w}\left[ \frac{1-\mu_w^2}{2} 
    \nu_{QL} (\mu_w)
    \frac{\partial f_w}{\partial \mu_w}\right], 
\end{equation}
which describes spatial advection along field lines as well as pitch angle diffusion through momentum space. Here, we introduced the quasilinear scattering rate $\nu_{QL}$ as well as the subscript ``$w$'' for quantities measured in the wave frame. Note that this equation is only valid in the wave frame since any other frame would require corrections to the RHS due to wave electric fields.

The quasilinear scattering rate is given by
\begin{equation} \label{eq:quasilinear_scattering}
    \nu_{QL} (p_w, \mu_w) = \pi \Omega k_{\mathrm{res}} I^{\pm} (k_{\mathrm{res}}),
\end{equation}
where the \alf wave power spectrum $I^{\pm}(k)$ is normalized through,
 \begin{equation} \label{eq:power_spectrum}
     \int I^{\pm} (k) \: dk = \left< \frac{\delta B_{\bot}^2}{B_0^2} \right>_x,
 \end{equation} 
and the notation $\langle ...\rangle_x$ indicates a spatial average. Above, we introduced the resonant wavenumber,
 \begin{equation} \label{eq:resonant_wavenumber}
     k_{\mathrm{res}} = \frac{m\Omega_c}{p_w \mu_w},
 \end{equation}
which corresponds to the wavenumber of an \alf wave in resonance with a particle with wave-frame momentum $p_w$ and pitch angle $\mu_w$. At $\mu_w = 0$, the resonant wavenumber is formally infinite, and no wave can scatter the particle. This situation is referred to as the 90$^{\circ}$ pitch angle problem (Felice \& Kulsrud 2001) and can be alleviated by resonance broadening, mirror scattering \citep{Holcomb2019}, or nonlinear wave-particle interactions \citepalias{Bai2019}. 

Wave-particle scattering relaxes the CR distribution to an isotropic distribution ``self-confined'' to co-move with \alf waves. Thus, Equation~\ref{eq:quasilinear_diffusion} is a specific instance of a Fokker-Planck equation for the pitch angle evolution of the distribution \citep{Shalchi2009},
\begin{equation} \label{eq:Fokker_Planck}
    \frac{\partial f_w}{\partial t} + 
    v_w \mu_w \frac{\partial f_w}{\partial x} = 
    \frac{\partial}{\partial \mu_w}\left[ D_{\mu \mu}
    \frac{\partial f_w}{\partial \mu_w}\right],
\end{equation}
where the Fokker-Planck (FP) diffusion coefficient is given by \cite{Mertsch2020} as
\begin{equation} \label{eq:Fokker_Planck_scattering}
    D_{\mu \mu} =
    \frac{\left< (\Delta \mu)^2 \right>}{2 \Delta t}.
\end{equation}
The angle brackets here represent an ensemble average (see Equations~\ref{eq:discrete_Fokker_Planck} and~\ref{eq:ensemble_average} for our numerical implementation of this calculation). 

By tracking individual particles, we can compute the change in pitch angle $\Delta \mu (t)$ for each particle and average over the distribution function to compute $D_{\mu \mu}$, which is related to the quasilinear scattering rate through
\begin{equation} \label{eq:FP_QL}
    D_{\mu \mu} (p_w,\mu_w) = \frac{1-\mu_w^2}{2} \nu_{QL} (p_w,\mu_w).
\end{equation}
Thus, the two rates presented here (Equations~\ref{eq:quasilinear_scattering} and~\ref{eq:Fokker_Planck_scattering}) provide independent means of computing the wave-particle scattering rate. \citetalias{Bai2019} previously showed that the distribution function evolves consistent with Equation~\ref{eq:quasilinear_diffusion} for particles away from $\mu$ = 0.   Here, we study individual particle orbits and compare their ensemble averaged scattering rates (via Equation~\ref{eq:Fokker_Planck_scattering}) to those predicted from quasilinear theory (via Equation~\ref{eq:quasilinear_scattering}). Since the Fokker-Planck rate captures nonlinear effects lost in the quasilinear approximation, comparison of these scattering rates provides insight into the strength of nonlinear versus quasilinear scattering.

\subsection{Fluid Theory}

Fluid theories reduce the dimensionality of statistical systems by averaging over moments of the underlying momentum space distribution. This procedure generates a hierarchy of moment equations, the truncation of which requires a closure. In our case, the closure will come from an assumption about the near-isotropy of the CR distribution. 

We define the particle energy as $\mathpzc{E}(p) = \gamma mc^2$. The CR energy density for a given momentum $\boldsymbol{p}_w$ is then 
\begin{equation} \label{eq:endensity_def}
    {\cal  E}_{\mathrm{CR},w} \equiv \int d\mu_w \: \mathpzc{E}({p}_w)
    f_w (\boldsymbol{p}_w),
\end{equation}
and the corresponding 
parallel CR energy flux in the wave frame is  
\begin{equation} \label{eq:flux_def}
    F_{\mathrm{CR},w} \equiv \int d \mu_w \:  \mathpzc{E}({p}_w)
v_{\parallel,w}(\boldsymbol{p_w}) 
    f_w (\boldsymbol{p}_w),
\end{equation}
where the parallel velocity $v_{\parallel,w}$~=~$v_w \mu_w$~=~$p_w \mu_w / (m\gamma_w)$. Note that formally, these quantities are energy density and flux \textit{per momentum density}. In Figures~\ref{fig:wave_energy_flux}-\ref{fig:space_time_fiducial}, we plot the total energy density and flux, i.e. the result of applying an additional integration $2\pi\int dp_w \: p_w^2$.

Multiplying the Fokker-Planck equation (\ref{eq:Fokker_Planck}) by the energy and energy flux per particle ($\mathpzc{E}({p})$ and $\mathpzc{E}({p}) v_{\parallel} (\boldsymbol{p})$ respectively) and integrating over $\mu_w$ and gyrophase $\phi$ yields
\begin{equation} \label{eq:continuity}
    \frac{\partial}{\partial t} {\cal E}_{\mathrm{CR},w} + \frac{\partial}{\partial x} F_{\mathrm{CR},w} = 0,
\end{equation}
\begin{equation} \label{eq:flux}
\begin{split}
    \frac{\partial}{\partial t} F_{\mathrm{CR},w} + \frac{\partial}{\partial x} \int d \mu_w \: v_w^2 \mathpzc{E} \mu_w^2 f_w =\\
    - \int d \mu_w \: \mathpzc{E} v_w 
D_{\mu \mu}
\frac{\partial f_w}{\partial \mu_w},
\end{split}
\end{equation}
where Equation~\ref{eq:continuity} represents conservation of CR energy density  and Equation~\ref{eq:flux} describes evolution of the CR flux. 
Because the momentum space integral is taken over pitch angle alone, these moment equations apply momentum- by-momentum.

Since $f_w$ is close to isotropic, we can assume that any anisotropy in the distribution function does not contribute substantially to the integral over $\mu_w$ for the second term on the left-hand side of Equation~\ref{eq:flux}. We therefore approximate the expression inside the gradient as 
\begin{equation}
    \int d \mu_w \: v_w^2 \mu_w^2 \mathpzc{E} f_w \approx 
    \frac{v_w^2}{3} \mathpzc{E}(p_w)\int d \mu_w \:  f_w = \frac{v_w^2}{3} {\cal E}
    _{\mathrm{CR},w}. 
\end{equation}
This assumption of approximate isotropy of the distribution relies upon $v_D/c \ll 1$, which is certainly true for relativistic CRs. Thus, CR hydrodynamics may still be applied for a distribution with rapid streaming as long as $v_A \ll v_D \ll c$.

We now introduce an effective, ``fluid'' scattering rate $\nu_{\mathrm{eff}}$, such that the pitch angle dependence of the distribution function and wave-particle scattering rate is entirely absorbed into $\nu_{\mathrm{eff}}$,
\begin{equation} \label{eq:fluid_equation}
\begin{split}
     \frac{\partial}{\partial t} F_{\mathrm{CR},w} + \frac{v_w^2}{3} \frac{\partial}{\partial x} {\cal E}_{\mathrm{CR},w} = -\nu_{\mathrm{eff}} F_{\mathrm{CR},w}.
\end{split}
\end{equation}
Here, 
\begin{equation} 
    \nu_{\mathrm{eff}} = 
    \frac{1}{F_{\mathrm{CR,w}}} 
    \int d \mu_w \:  \mathpzc{E} v_w 
    D_{\mu \mu} \frac{\partial f_w}{\partial \mu_w},
\end{equation}
and for quasilinear theory substitution of Equation~\ref{eq:FP_QL} results in 
\begin{equation} \label{eq:effective_scattering1}
    \nu_{\mathrm{eff,QL}} = 
    \frac{1}{F_{\mathrm{CR,w}}} 
    \int d \mu_w \: \mathpzc{E} v_w 
    \left( \frac{1-\mu_w^2}{2} \right) \nu_{QL} \frac{\partial f_w}{\partial \mu_w}.
\end{equation}

For a flat spectrum with $k_{\mathrm{res}} I^{\pm} (k_{\mathrm{res}})$ = constant, Equation~\ref{eq:effective_scattering1} reduces to $\nu_{\mathrm{eff,QL}} = \nu_{QL}$. We treat the flux moment equation (\ref{eq:fluid_equation}) as a function of momentum, computing the effective scattering rate for each momentum bin. This procedure allows us to compare the quasilinear, Fokker-Planck, and fluid scattering rates as a function of particle momentum.

\section{Methods} \label{sec:methods}

We wish to study the spatial and temporal evolution of the CR distribution function and compare this evolution to predictions from fluid theory. This evolution arises naturally at the interface of two transport regimes: an ambient ISM where particle transport is diffusive and a cloud where \alf waves are damped through ion-neutral collisions and CRs stream freely. 
\subsection{The MHD-PIC Method}

The magnetohydrodynamic particle-in-cell (MHD-PIC) method is a plasma model which evolves a kinetic species (CRs) under the influence of force fields calculated using the equations of MHD \citep{Bai2015}. For the gyroresonant CRSI, the thermal plasma, referred to as the ``gas'' with the subscript ``g'' is described by the equations of ideal MHD with source terms:
\begin{equation} \label{eq:mass_conservation}
    \frac{\partial \rho}{\partial t} + \nabla \cdot \left( \rho \boldsymbol{v}_g \right) = 0,
\end{equation}
\begin{equation} \label{eq:momentum_conservation}
\begin{split}
    \frac{\partial \left( \rho \boldsymbol{v}_g \right)}{\partial t} + \nabla \cdot \left( \rho \boldsymbol{v}_g \boldsymbol{v}_g - \boldsymbol{\boldsymbol{B}} \boldsymbol{\boldsymbol{B}} + P_{\mathrm{Tot}} \right) \\
    = - \left(q n_{\mathrm{CR}} 
    \boldsymbol{E} + \frac{\boldsymbol{J}_{\mathrm{CR}}}{c} \times \boldsymbol{B} \right) - \nu_{\mathrm{IN}} \rho \boldsymbol{v}_{\bot g},
\end{split}
\end{equation}
\begin{equation} \label{eq:energy_conservation}
    \frac{\partial {\cal E}_{\mathrm{Tot}}}{\partial t} + \nabla \cdot \left[ \left( {\cal E}_{\mathrm{Tot}} + P_{\mathrm{Tot}} \right) \boldsymbol{v}_g - \left( \boldsymbol{{B}} \cdot \boldsymbol{v}_g \right) \boldsymbol{{B}} \right] = - \boldsymbol{J}_{\mathrm{CR}} \cdot \boldsymbol{E}.
\end{equation}
Here, $\rho$ is the gas density, $\mathcal{I}$ is the unit tensor, the total pressure $P_{\mathrm{Tot}}$~=~$P_g$~+~$B^2/2$, and the electric field $\boldsymbol{E}$~=~$-\boldsymbol{v}_g \times \boldsymbol{B}/c$. The total energy is given by
\begin{equation}
    {\cal{E}}_{\mathrm{Tot}} = \frac{P_g}{\gamma_{\mathrm{ad}} - 1} + \frac{1}{2} \rho v_g^2 + \frac{1}{2} |\boldsymbol{B}|^2,
\end{equation}
where $\gamma_{\mathrm{ad}}$ is the adiabatic index. We work in units where the magnetic permeability is unity such that $4 \pi$~=~1.

Ion-neutral damping is implemented as a simple exponential attenuation factor,
\begin{equation}
    \boldsymbol{v}_{\bot g} \rightarrow 
    \boldsymbol{v}_{\bot g} e^{-\Gamma_{\mathrm{IN}} \Delta t},
\end{equation}
where $\Delta t$ is the simulation time step and $\Gamma_{\mathrm{IN}}$ is the ion-neutral damping rate. The details of this damping including numerical properties and the effect on the growth and saturation of the gyroresonant CRSI is discussed in detail by Plotnikov et al. (2021 in prep.).

Particles are pushed by the Lorentz force,
\begin{equation}
    \frac{d \boldsymbol{p}_j}{dt} = \left( \frac{q}{m c} \right)_j \left(c \boldsymbol{E} + \boldsymbol{v}_j \times \boldsymbol{B} \right),
\end{equation}
where ``j'' refers to the $j$th particle, $\boldsymbol{p_j} = \gamma_j \boldsymbol{v}_j$, and the charge to mass ratio $q/mc$ $\equiv$ 1. 

Since we only use 16 particles per cell per type, Poisson noise due to particle discreteness is a major limitation. We compensate for this noise by using a $\delta f$-method \citep{Dimits1993,Parker1993,Hu1994,Denton1995,Kunz2014_Pegasus,Bai2019}.
In the $\delta f$-method, the distribution function $f(t,x,\boldsymbol{p})$ is split into a uniform static background $f_0 (\boldsymbol{p})$ and the particles evolved in the simulation, represented by a perturbation on this background, $\delta f (t,x,\boldsymbol{p})$. The PIC method itself pushes particles in phase space in order to compute a weighting function for the jth particle, $w_j$:
\begin{equation}
    w_j = \frac{\delta f(t,x_j (t), \boldsymbol{p}_j (t))}{f(t,x_j (t), \boldsymbol{p}_j (t))} =  1 - \frac{f_0(x_j(t), \boldsymbol{p}_j(t))}{f(0,x_j(0),\boldsymbol{p}_j(0))}.
\end{equation}
We can then straightforwardly compute the CR number and current densities in Equations~\ref{eq:momentum_conservation} and~\ref{eq:energy_conservation},
\begin{equation}
\begin{split}
    n_{\mathrm{CR}} = n_{\mathrm{CR},0} + \int \delta f(t,x,\boldsymbol{p}) \: d^3 \boldsymbol{p} \\ \simeq n_{\mathrm{CR},0} + \sum_{j=1}^{N_p} w_j S(x-x_j),
\end{split}
\end{equation}
\begin{equation}
\begin{split}
    \boldsymbol{J}_{\mathrm{CR}} (t,x) = \boldsymbol{J}_{\mathrm{CR},0} + q_{\mathrm{CR}} \int \boldsymbol{v} \delta f(t,x,\boldsymbol{p}) \: d^3 \boldsymbol{p} \\ \simeq \boldsymbol{J}_{\mathrm{CR},0} + q_{\mathrm{CR}} \sum_{j=1}^{N_p} w_j \boldsymbol{v}_j S(x-x_j),
\end{split}
\end{equation}
where the summations are over the total number of particles $N_p$, the subscript ``0'' represents moments of the background distribution $f_0 (\boldsymbol{p})$, and $S(x-x_j)$ is the shaping function for interpolating a point particle at $x$-coordinate $x_j$ to the grid. We use a triangular-shaped cloud \citep[TSC;][]{Birdsall2005} for $S(x-x_j)$.

The simulations are run in a one-dimensional box with periodic boundaries and an equilibrium magnetic field $\boldsymbol{B}_0 = B_0 \hat{x}$. Particles are limited by the numerical speed of light $\mathbbm{C}$ = 300 $v_A$, which reduces the separation between wave velocities ($v_A$) and particle velocities ($\sim$~$\mathbbm{C}$). Any analysis involving the speed of light $c$ (computation of energy density or flux), uses this numerical speed of light $c$ = $\mathbbm{C}$. Similarly, Lorentz transformations use this numerical speed of light. Since the error in these transformations scales as $\mathcal{O} (v_A/\mathbbm{C})$, our choice of $\mathbbm{C}$ ensures $v_A/\mathbbm{C} \ll 1$ while reducing computational cost. 

Particles are evolved via the energy-conserving Boris pusher \citep{Boris1970} and the MHD-PIC equations (\ref{eq:mass_conservation}-\ref{eq:energy_conservation}) are solved using the \texttt{Athena} MHD code \citep{Stone2008}, with constrained transport to ensure $\nabla \cdot \boldsymbol{{B}}$~=~0 \citep{Evans1988}. Time integration is done through the corner transport upwind method \citep{Gardiner2005, Gardiner2008} and we use the Roe Riemann solver \citep{Roe1981} with third-order reconstruction for spatial integration. The details of the MHD-PIC method can be found in \cite{Bai2015} and \citetalias{Bai2019}. 

\subsection{Modifications to Previous Simulations}

\begin{table*}
\caption{Summary of Simulations}  
\label{table:sims} 
\renewcommand{\arraystretch}{1.1}
\small\addtolength{\tabcolsep}{-2pt}
\begin{center}
\scalebox{1}{%
\begin{tabular}{c c c c c c}     
\hline  
Simulation & $n_{\mathrm{CR,0}}/n_i$ & $v_D/v_A$ &     $\nu_{\mathrm{IN}}$ & $L_{\mathrm{ISM}}/L_{\mathrm{Cloud}}$ & $L$ \\ 
           &                         &           &  
($\Omega_c$)        &                & ($v_A \Omega_c^{-1}$)\\      
\hline
{{Fiducial}}& 10$^{-4}$ & 10 & 5.08$\times$10$^{-4}$ & 1/1 & 2$\times$10$^6$ \\
{{NCR2}}& 2$\times$10$^{-4}$ & 10 & 10.16$\times$10$^{-4}$ & 1/1 & 2$\times$10$^6$ \\
{{Long\_Cloud}}& 10$^{-4}$ & 10 & 5.08$\times$10$^{-4}$ & 1/4 & 5$\times$10$^6$\\
{{Long\_ISM}}& 10$^{-4}$ & 10 & 5.08$\times$10$^{-4}$ & 4/1 & 5$\times$10$^6$\\
\hline
\end{tabular}}
\\
\end{center}
\end{table*}

\citetalias{Bai2019} studied a suite of magnetohydrodynamic particle-in-cell (MHD-PIC) simulations of the gyroresonant CR streaming instability (CRSI) which varied the equilibrium number density ($n_{\mathrm{CR,0}}$) and drift velocity ($v_D$) to study the growth  of the instability and quasilinear evolution of the particle distribution. Their simulations were performed in a frame co-moving with the CR drift velocity such that the momentum space distribution of the particles was described by an isotropic $\kappa$ distribution \citep{Summers1991},
\begin{equation} \label{eq:kappa}
    f_{\kappa}(p) = \frac{n_{\mathrm{CR,0}}}{\left(\pi \kappa p_0^2 \right)^{3/2}} \frac{\Gamma(\kappa+1)}{\Gamma(\kappa-\frac{1}{2})} \left[ 1 + \frac{1}{\kappa} \left(\frac{p}{p_0} \right)^2 \right]^{-(\kappa+1)},
\end{equation}
where $p_0$ = 300 $v_A$ is the peak momentum of the distribution and $\kappa$ = 1.25. Waves in the \citetalias{Bai2019} simulations were able to grow unimpeded by physical damping mechanisms, succumbing only to numerical diffusion. 

The cloud problem investigated in this paper requires three modifications to the \citetalias{Bai2019} simulations:
\begin{enumerate}
    \item Rest Frame rather than Drift Frame: For the cloud to be stationary within the simulation (lab) frame, the simulation must be performed in the rest frame of the thermal plasma rather than in a frame drifting with the CRs. Because particles are relativistic, transformations of the $\kappa$ distribution to this new frame require a Lorentz transformation.
    \item Large Box Size: Particles are allowed to traverse $\sim$10 mean free paths to scattering with waves within the ambient ISM region (hereafter ``ISM''). The cloud is made comparable in size to the ISM to capture large-scale evolution. 
    \item Ion-neutral Damping of Waves: \alf waves are attenuated in the cloud region while being allowed to grow unimpeded in the ISM.
\end{enumerate}

Related to (1), we have verified that we can recover the \citetalias{Bai2019} results, including wave growth rates, power spectra, and distribution functions for simulations with no wave damping independent of frame.
Related to (2), based on calculations of mean free paths in runs of the \citetalias{Bai2019} problem with varying box size (see Appendix~\ref{sec:box_size}), we choose the minimum ISM length,
\begin{equation} \label{eq:ISM_length}
    L_{\mathrm{ISM}} = 10^6 v_A \Omega_c^{-1},
\end{equation}
which corresponds to $\approx 4$  mean free paths in the ISM region for $n_{\mathrm{CR,0}}$/$n_i$ = 10$^{-4}$. For higher CR number densities, more mean free paths are present in the ISM. 

Table~\ref{table:sims} summarizes parameters for the 4 simulations presented in this work. 
For all simulations in this paper, we use a constant initial drift velocity $v_D/v_A$~=~10 to increase the wave growth rate, saturation amplitude, and initial CR flux relative to the fiducial model of \citetalias{Bai2019}.  
The \texttt{Fiducial} simulation, where the ISM length $L_{\mathrm{ISM}}$ and the cloud length $L_{\mathrm{Cloud}}$ are equal and $n_{\mathrm{CR,0}}/n_i$~=~$10^{-4}$, is our primary focus. \texttt{NCR2} studies the case where the waves grow more rapidly and the scattering rate is twice that in the \texttt{Fiducial} run. By extending the cloud region in the simulation \texttt{Long\_Cloud}, the CR energy flux decreases more slowly and becomes a subdominant term in the fluid equation. Finally, by extending the ISM region relative to the cloud, the \texttt{Long\_ISM} simulation studies the situation where the CR distribution is isotropized more rapidly. All simulations are computed in a box at least 20 times longer than that studied in \citetalias{Bai2019}, with each grid cell spanning 10 $v_A \Omega_c^{-1}$, where we work in units of $v_A$~=~$\Omega_c$~=~1. For the \texttt{Long\_Cloud} and \texttt{Long\_ISM} simulations, the box is 50 times longer. Thus, all simulations presented are computed with only 16 particles per cell per type (where 8 different particle types are used which span the full range of momentum, $-2 \leq \log{(p_d)} \leq 2$) until $t$~=~10$^5$~$v_A \Omega_c^{-1}$.

\subsubsection{Rest Frame vs. Drift Frame} \label{sec:frame_transformation}

\citetalias{Bai2019} did their calculations in the initial drift frame of the CRs, implying that their background thermal plasma initially had a velocity $-v_D \hat x$. We shall instead work in the frame where the initial background fluid (both ISM and cloud) are stationary, so that the CR distribution is isotropic in a frame moving with velocity $v_D \hat x$.  We shall refer to the initial rest frame of the background plasma (the simulation frame) as either the ``rest'' or ``lab'' frame, and the frame where the initial CR distribution is isotropic as the ``drift'' frame.  

 The 4-momentum of a particle $p^{\mu}$ is defined as  
\begin{equation}
    p^{\mu} = \gamma (v) \: \left(c,v_x,v_y,v_z \right),
\end{equation}
where $\gamma$ is the Lorentz factor (Equation~\ref{eq:gamma}). The magnitude of a 4-vector, $p^{\mu} p_{\mu}$, is invariant under a Lorentz boost along the $x$-direction, $\boldsymbol{\Lambda}_{\: \: \nu}^{\mu}$:
\begin{equation}
    \boldsymbol{\Lambda}_{\: \: \nu}^{\mu} 
    \left[ v_D \right] = 
    \begin{pmatrix} 
     \gamma (v_D) & -\frac{v_D}{c} \gamma (v_D) & 0 & 0 \\ 
     -\frac{v_D}{c} \gamma (v_D) & \gamma (v_D) & 0 & 0 \\
            0                    &        0     & 1 & 0 \\
            0                    &        0     & 0 & 1
    \end{pmatrix},
\end{equation}
\begin{equation}
    p^{\mu}_{\mathrm{drift}} = 
    \boldsymbol{\Lambda}_{\: \: \nu}^{\mu} \left[ v_D \right]
    p^{\nu}_{\mathrm{rest}}.
\end{equation}
The inverse of $\boldsymbol{\Lambda}_{\: \: \nu}^{\mu}$, needed to boost from drift frame to lab frame, is trivially computed with the substitution $v_D \rightarrow -v_D$.

In initializing the particles, we employ the $\kappa$ distribution, but because this applies in the drift frame we must boost to the lab frame.  In the lab frame, the CR distribution is not initially isotropic; the anisotropy results purely from the drift velocity of the distribution. The transformation of the $\kappa$ distribution $f_{\kappa} (p_{\mathrm{drift}})$ is then
\begin{equation}
    f_{\kappa} (p_{\mathrm{drift}}) = f_{\kappa} 
    \left( \left| \boldsymbol{\Lambda}_{\: \: \nu}^{\mu} \left[ v_D \right] p^{\nu}_{\mathrm{rest}} \right| \right) = f_0 (p_{\mathrm{rest}}),
\end{equation}
where the underlying distribution is unchanged as Lorentz transformations preserve phase space volume. Here, the magnitude symbol refers to magnitude of the 3-momentum, i.e. $\sqrt{p_1^2 + p_2^2 + p_3^2}$. Because of conservation of particle number, the Lorentz boost transforms the isotropic $\kappa$ distribution into a drifting prolate distribution unstable to the gyroresonant CRSI.

In the isotropic frame, the distribution has no net velocity, so there is no net current; however, in the rest frame, the net drift of the distribution creates significant current in the $x$-direction. This current is computed by boosting the 4-current density $J^{\mu}_{\mathrm{CR}}$ = $(q_{\mathrm{CR}} n_{\mathrm{CR,0}}, \boldsymbol{J}_{\mathrm{CR,0}})$ from the drift frame of the CRs to the gas rest frame.

\subsubsection{CRSI Growth Rate \& Ion-Neutral Damping}

The atomic and molecular ISM has ionization fractions ranging from $\sim 10^{-2}$ in the warm atomic gas to $\sim 10^{-4}$ in the cold atomic gas to $\lesssim 10^{-6}$ in the molecular gas \citep{Draine2011}.
Many neutral atoms and molecules are present, serving as targets for the few ions carrying \alf waves. These neutrals provide a collisional drag force on the ions, which removes their momentum and damps the waves. 

In this work, we parameterize the ion-neutral damping rate based on the peak growth rate of the gyroresonant CRSI in the absence of damping. The growth rate as a function of $k$ can be worked out from the full Vlasov-Maxwell system with a drifting population of CRs \citep{Zweibel2003,Amato2009}. Using the notation introduced in \citetalias{Bai2019}, the growth rate is
\begin{equation} \label{eq:growth_rate}
    \Gamma_{\mathrm{CR}} (k) = \frac{1}{2} \frac{n_{\mathrm{CR,0}}}{n_i} \Omega_c \left( 
    \frac{v_D}{v_A} - 1
    \right) Q_2 (k),
\end{equation}
where for a $\kappa$ distribution,
\begin{equation}
    Q_2 (k) = \frac{\sqrt{\pi}}{\kappa^{3/2}} \frac{\Gamma(\kappa + 1)}{\Gamma(\kappa - \frac{1}{2})} \frac{1}{s_0 \left[1+ 1/(\kappa s_0^2) \right]^{\kappa}},
\end{equation}
and $s_0 = k p_0/(m \Omega_c)$. The growth rate is maximized at $s_0 = \sqrt{2-1/\kappa}$. For $\kappa$ = 1.25, $p_0$~=~300~$v_A$, and $v_D$~=~10~$v_A$ used throughout this work, the peak growth rate is $\Gamma_{\mathrm{CR}}(k_{\mathrm{peak}})$~$\approx$~2.54$(n_{\mathrm{CR,0}}/n_i)$~$\Omega_c$. 

In low-ionization regions, the damping rate of \alf waves depends in general on the wave frequency $\omega$. For trans-relativistic CRs and typical conditions in atomic and molecular gas,  $\omega = k v_{A,i} \gg \nu_{\mathrm{IN}}$, where $v_{A,i}$ is the \alf speed considering just the ions\footnote{For convenience, in general we omit the ``i'' subscript on the \alf speed, but here we include it to emphasize that the relevant \alf speed is $v_{A,i} = B/\rho_i^{1/2}$.}, and $\nu_{\mathrm{IN}}$ is the ion-neutral collision rate (see Plotnikov et al. 2021, in prep. for detailed discussion of ISM conditions). In this limit, the  damping rate of waves is simply
\begin{equation}
    \Gamma_{\mathrm{IN}} = \frac{\nu_{\mathrm{IN}}}{2}.
\end{equation}
We define the critical collision frequency as the rate where the peak wave growth equals damping,
\begin{equation} \label{eq:critical_damping}
    \nu_{\mathrm{crit}} = 2\Gamma_{\mathrm{CR}}(k_{\mathrm{peak}}) = 5.08 \left( \frac{n_{\mathrm{CR,0}}}{n_i} \right) \: \Omega_c.
\end{equation}

Throughout this work, we set the ion-neutral damping rate (in the cloud) to be this critical rate. We performed an exploration of various damping rates, concluding that any damping rate above the critical rate (even 10$^3$ times this rate) produces the same particle evolution as the $\nu_{\mathrm{IN}} = \nu_{\mathrm{crit}}$ case. This is perhaps not surprising, since any damping rate above $\nu_{\mathrm{crit}}$ will completely suppress wave growth, resulting in ballistic propagation through the cloud. Plotnikov et al (2021, in prep.) consider simulations with a range of $\nu_{\rm IN}/\nu_{\rm crit} = 0.015 - 1 $.

\section{Emergence of Spatial Structure} \label{sec:spatial_structure}

We are studying a time-dependent problem wherein a distribution of cosmic rays (CRs) drives the growth of \alf waves through the gyroresonant CRSI. Simultaneously, wave-particle scattering slowly removes the anisotropy of the drifting CR distribution, decreasing the drift velocity $v_D$. Because CRs diffuse through the ISM region and free stream through the cloud, spatial structure inexorably emerges in the distribution function. A negative spatial gradient in the CR energy density spans the ISM while a positive gradient spans the cloud. The minimum in the CR energy density is located at $x$ = 10$^6$ $v_A \Omega_c^{-1}$ in the \texttt{Fiducial} simulation, which we define as the ``leading ISM-cloud interface.''

In this section, we explore the origin of this energy density gradient, which appears universally in our cloud simulations. First, we introduce the diagnostics used to study the CRs and waves.

The \alf wave energy density $\mathcal{E}_A$ is defined,
\begin{equation}
    \mathcal{E}_A = \frac{1}{2} \rho \left| \delta \mathbf{v}_{\bot} \right|^2 + 
    \frac{\left| \delta \mathbf{B}_{\bot} \right|^2}{2} = \left| \delta \mathbf{B}_{\bot} \right|^2,
\end{equation}
where $\mathbf{v}_{\bot} = v_y \hat{y} + v_z \hat{z}$ and $\mathbf{B}_{\bot} = B_y \hat{y} + B_z \hat{z}$ are the perpendicular components of the velocity and magnetic fields respectively. Using the $\delta f$-method formalism, we can define the CR number and energy densities and the parallel energy flux at the $i$th position $x_i$,
\begin{equation} \label{eq:number_density}
\begin{split}
    n_{\mathrm{CR}} (t,x_i) &= \int d^3 \boldsymbol{p} \: \left[ f_0(\boldsymbol{p}) + \delta f(t,x_i,\boldsymbol{p})\right],
\end{split}
\end{equation}
\begin{equation} \label{eq:energy_density}
\begin{split}
    \mathcal{E}_{\mathrm{CR}} (t,x_i) &= \int d^3 \boldsymbol{p} \: \mathpzc{E}(p) \left[f_0(\boldsymbol{p}) + \delta f(t,x_i,\boldsymbol{p}) \right],
\end{split}
\end{equation}
\begin{equation} \label{eq:energy_flux}
\begin{split}
    F_{\mathrm{CR}} (t,x_i) &= \int d^3 \boldsymbol{p} \: v_{\parallel} (\boldsymbol{p}) \mathpzc{E}(\boldsymbol{p}) \left[f_0(\boldsymbol{p}) + \delta f(t,x_i,\boldsymbol{p}) \right].
\end{split}
\end{equation}
We can convert between energy densities and fluxes in different frames by constructing the energy flux 4-vector in the rest frame of the CRs,
\begin{equation}
   F^{\mu}_{\mathrm{rest}} = \left({\cal E}_{\mathrm{CR}} c, F_{\mathrm{CR}}, 0,0 \right),
\end{equation}
and boosting into the frame of the waves,
\begin{equation}
    F^{\mu}_{\mathrm{wave}} = \boldsymbol{\Lambda}_{\: \: \nu}^{\mu} \left[v_A \right] F^{\nu}_{\mathrm{rest}}.
\end{equation}
This procedure can be used to find the background flux and energy density,
\begin{equation}
    {\cal E}_{\mathrm{CR,0}} = 3.62 \times 10^5 \: n_{\mathrm{CR,0}} \:\: \rho_0 v_A^2,
\end{equation}
\begin{equation}
    F_{\mathrm{CR,0}} = 3.62 \times 10^5 \: n_{\mathrm{CR,0}} \: v_D \:\: \rho_0 v_A^2,
\end{equation}
where we work in units of the background gas density $\rho_0$ and \alf velocity $v_A$. 

\begin{figure*}[!htb]
\hbox{
\psfig{figure=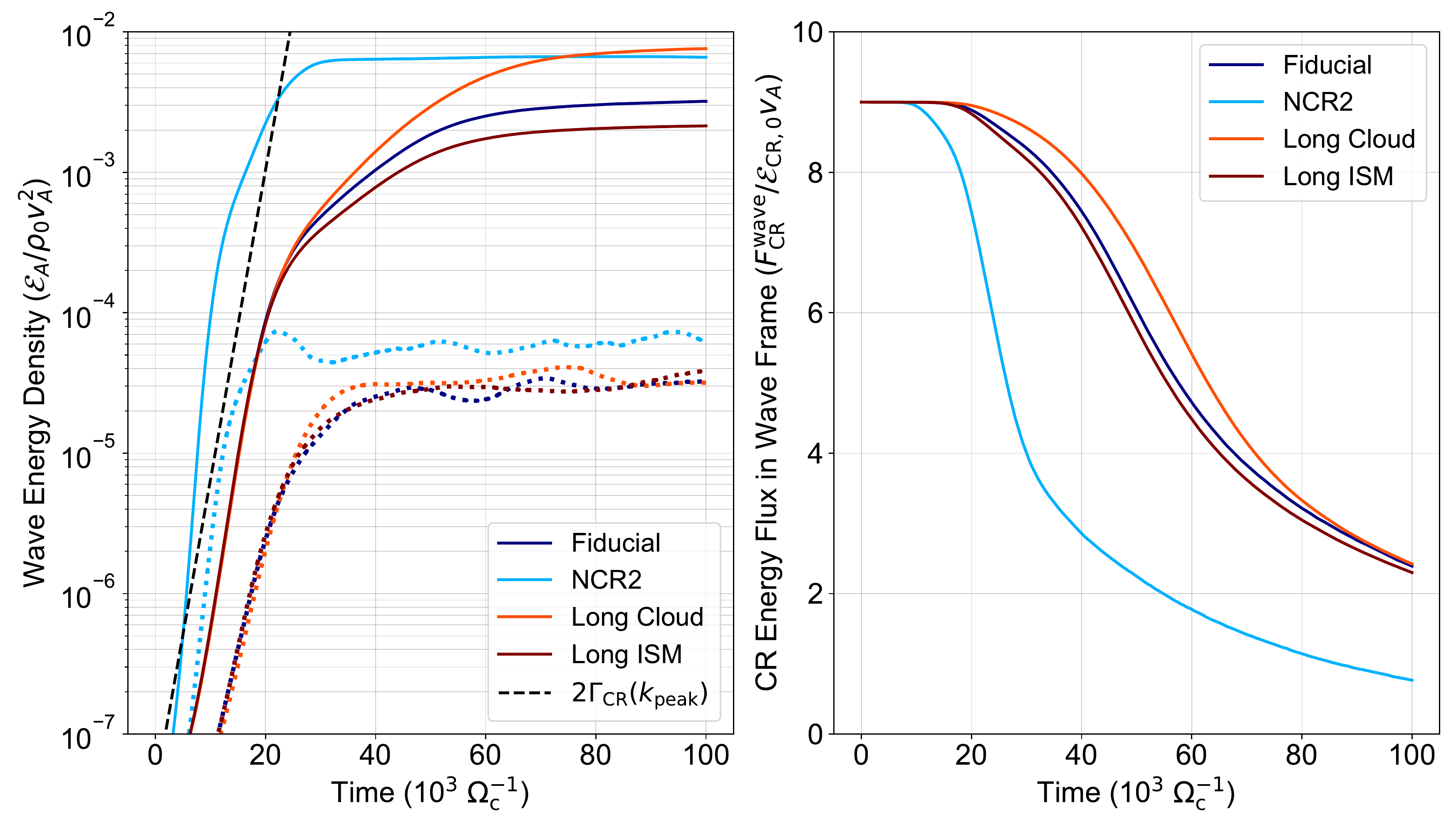,width=1.0\textwidth}
}
\caption{{\it Left}: \alf wave energy densities $\mathcal{E}_A$ in ISM (solid) and cloud (dotted) regions respectively for all simulations. In the ISM region, all simulations show growth rates consistent with the theoretical expectation, $2 \Gamma_{\mathrm{CR}} (k_{\mathrm{peak}})$ (black dashed line).  The saturation amplitudes vary substantially depending on the ISM to cloud length ratio and $n_{\mathrm{CR},0}/n_i$.  {\it Right}: Flux of CRs, as measured in the wave frame, in units $\mathcal{E}_{\mathrm{CR},0}v_A$.  For all simulations, the flux declines in time as the CR distribution is isotropized by particle-wave scattering.}
\label{fig:wave_energy_flux}
\end{figure*}

The temporal evolution of the \alf wave energy density and CR flux as measured in the wave frame is shown in Figures~\ref{fig:wave_energy_flux}a and~\ref{fig:wave_energy_flux}b respectively. While waves grow at the rate predicted by Equation~\ref{eq:growth_rate}, the saturation amplitudes both in the ISM and cloud can differ from the \texttt{Fiducial} run by up to a factor of 4. This is not surprising since the \texttt{Long\_Cloud} simulation has 2.5$\times$ the number of particles as the \texttt{Fiducial} run and thus 2.5$\times$ the CR momentum. Particle scattering transfers this momentum to \alf waves \citep{Kulsrud2005,Bai2019}; however, strongly-suppressed waves in the cloud are unable to grow sufficiently to scatter and receive momentum at a significant rate. Nearly all momentum lost by the CRs must be transferred to waves in the ISM. Similarly, the saturation amplitude in the \texttt{Long\_ISM} simulation is lower than that of the \texttt{Fiducial} run since the ISM length is increased by a factor of 4 while the total CR momentum to be deposited increases by only a factor of 2.5.

\subsection{Wave and Cosmic Ray Co-evolution}

\begin{figure*}[!htb]
\hbox{
\psfig{figure=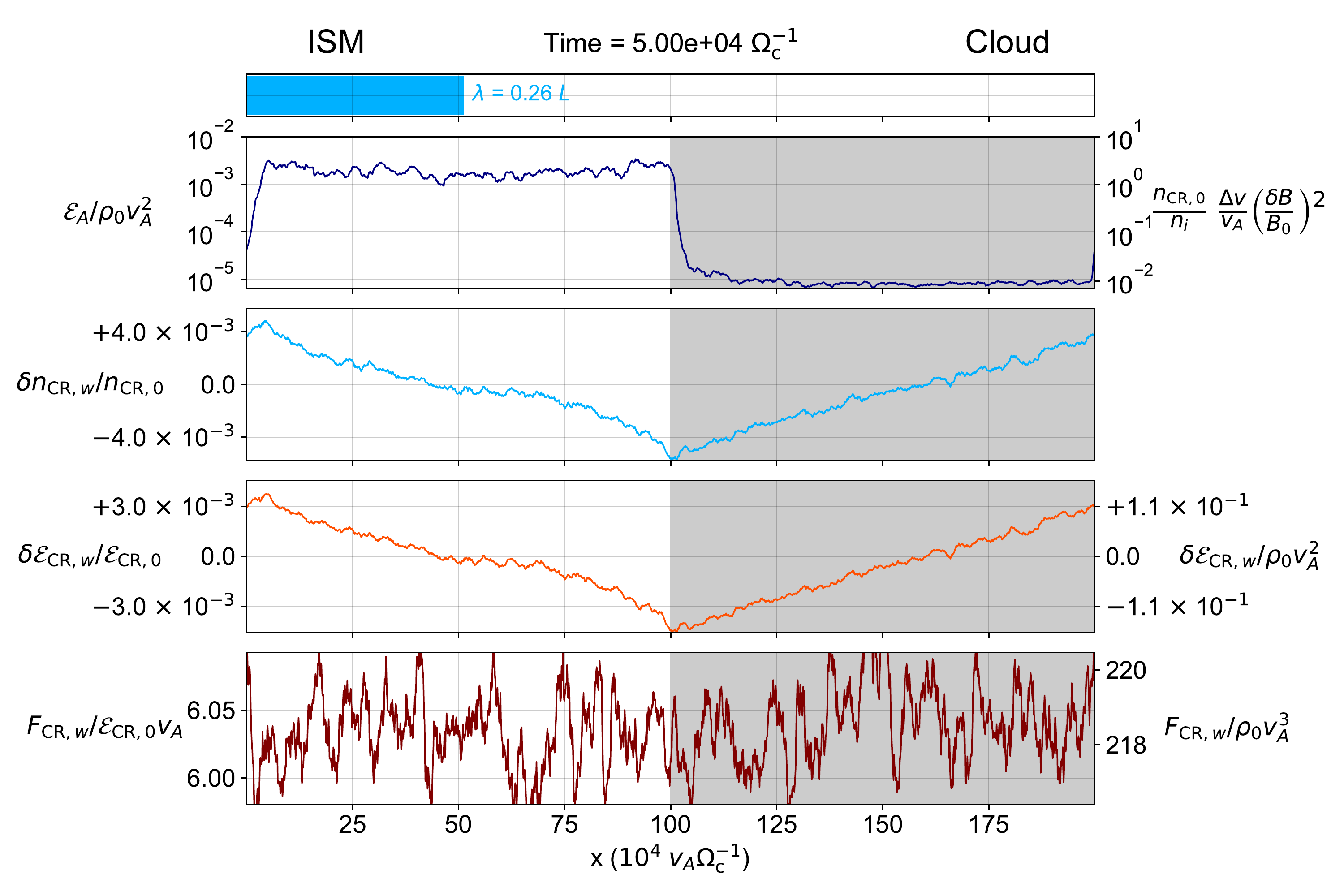,width=1.0\textwidth}
}
\caption{Spatial distribution of wave and particle energy and flux for \texttt{Fiducial} simulation, shown  at time $t$~=~5$\times$10$^4$~$v_A \Omega_c^{-1}$. 
All quantities are smoothed using convolution with a smoothing length of 0.01$L$. The mean free path for the ISM is shown (light blue bar), and the cloud region is shaded. At this time, the energy density and number density of the CRs form a negative gradient across the ISM region and a positive gradient across the cloud.  The gradient develops in the ISM due to spatial diffusion as CRs are scattered by waves (top row). The gradient in the cloud is a consequence of the periodic boundary conditions which must balance the CR pressure across the ISM-cloud interface. The profile is consistent with predictions from CR fluid theory in a periodic domain.
}
\label{fig:fiducial_gradient}
\end{figure*}

For all simulations, the average wave-frame energy flux decreases with time; however, no simulation reaches a fully isotropic state within the simulation time. Because waves grow most rapidly in the \texttt{NCR2} simulation and grow to an amplitude larger than all but the \texttt{Long\_Cloud} run, scattering is frequent and the distribution is more rapidly evolved toward isotropy.  The \texttt{Long\_Cloud} simulation isotropizes slowly since at any given time, the majority of particles are located in the cloud region and experience no scattering. 
In Figure~\ref{fig:fiducial_gradient}, we show the profiles of the \alf wave amplitude together with  the CR number density, energy density, and flux, at time $5\times 10^4 \Omega_c^{-1}$. At this stage,  Figure~\ref{fig:wave_energy_flux} shows that the wave energy is saturated and the CR flux has significantly declined from its initial value. Figure~\ref{fig:fiducial_gradient} conveys one of the main results of this paper: the presence of a cloud region allows for the formation of a spatial gradient in the CR number and energy densities.  The maximum is on the upstream side of the ISM (near $x = 0$) while the minimum is where the streaming CRs enter the cloud from the ISM (near $x=10^5 v_A \Omega_c^{-1}$ for the fiducial model).   

The gradients in the profiles of ${\cal E}_{\rm CR}$ and $n_{\rm CR}$ can be understood within the context of the CR fluid flux equation, Equation~\ref{eq:fluid_equation}. In the cloud, \alf wave growth is suppressed and the scattering rate remains small. Thus, the RHS of Equation~\ref{eq:fluid_equation} can be ignored. Since the energy flux is always decreasing in time (Figure~\ref{fig:wave_energy_flux}b), a positive spatial energy density gradient must form across the cloud to balance the temporal drop in flux. 

In the ISM, a negative gradient forms from particle diffusion as CRs traverse the region. Since the simulation is periodic, the CR energy density (pressure) gradients must balance one another on the upstream side of the ISM (which is also the downstream side of the cloud) located at the domain boundary. The temporal decrease in flux and negative spatial energy density gradient must work together to balance the negative scattering rate term on the right-hand side of Equation~\ref{eq:fluid_equation}. 
As we shall show, the magnitude of the negative number/energy density gradient is consistent with fluid theory for the measured CR flux based on the predicted effective scattering rate. 

Since our analysis relies on constructing an ISM where wave-particle scattering is sufficient for particles to propagate diffusively, we verify that multiple mean free paths are present in the ISM upon saturation. The mean free path to scattering $\lambda_{\mathrm{mfp}}$ for a particle with momentum coordinates ($p$,$\mu$) can be estimated as
\begin{equation} \label{eq:mfp}
    \lambda_{\mathrm{mfp}} (x) = \int_{0}^{1} d \mu \: \frac{|\mu| v}{\nu_{\mathrm{scat}}} \sim \frac{4}{\pi}
    \frac{p}{\Omega_c} 
    \left( \frac{\delta B_{\bot} (x)}{B_0} \right)^{-2},
\end{equation}
where we have used the approximate scattering rate from Equation~\ref{eq:naive_scattering}. We set $p$ = $p_0$ above since we are most interested in the peak of the distribution where the majority of particles are present. In the top panel of Figure~\ref{fig:fiducial_gradient}, we show the mean free path in the ISM, $\lambda = 0.26 L$, based on wave amplitudes averaged over this region. The mean free path in the cloud is $\approx$18.2$L$; particles free stream with no significant wave-particle collisions.

At the time $t$ = 5$\times$10$^4$ $v_A \Omega_c^{-1}$ in the \texttt{Fiducial} simulation, particles traverse approximately 2 mean free paths in the ISM. This mean free path is not particularly short, perhaps calling into question the validity of a diffusion approximation for particle transport. Based on tracking particle trajectories, we find that indeed even though CRs undergo many small angle scattering events, they experience few direction reversals over the simulation time. Yet, gradient structures are nonetheless able to develop in the ISM and cloud. Thus, while particle propagation may not be strictly diffusive in real space, exhibiting a random walk in $x$, CRs in our simulations evolve diffusively in momentum space. In this way, even 2 mean free paths in our simulations can roughly capture the diffusion process. 
\begin{figure*} [!htb]
\hbox{
\psfig{figure=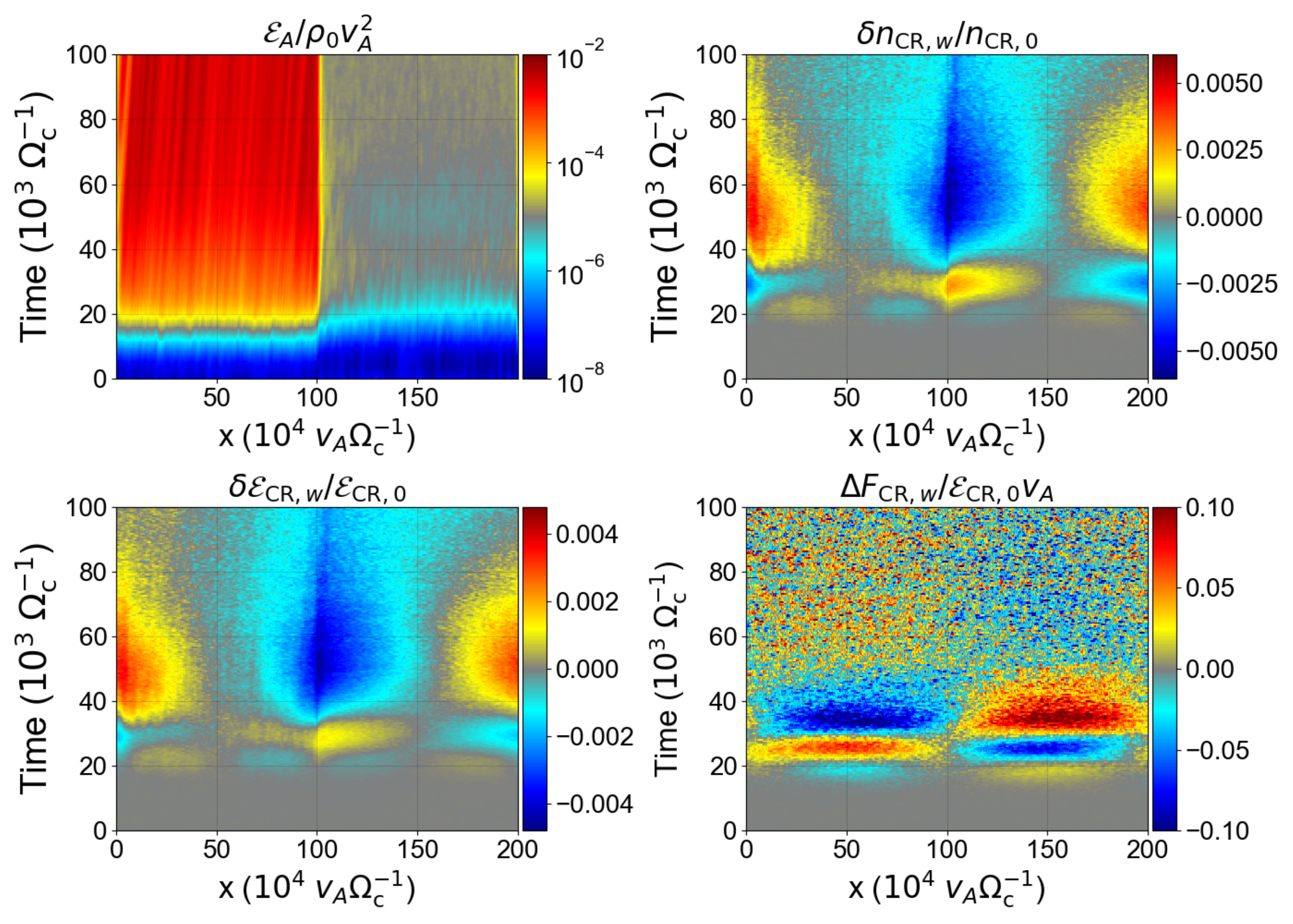,width=1.0\textwidth}
}
\caption{Space-time evolution of wave energy density $\mathcal{E}_A$,  perturbed particle number $\delta n_{\rm CR,w}/n_{\rm CR,0}$ and energy $\delta \mathcal{E}_{\rm CR,w}/\mathcal{E}_{\rm CR,0}$ densities, and flux $\Delta F_{\mathrm{CR},w}$. The fluctuation in the CR energy flux is defined $\Delta F_{\mathrm{CR},w} (t) \equiv F_{\mathrm{CR},w} (t) - \langle F_{\mathrm{CR},w} (t) \rangle_x$, where the angle brackets denote an average over the simulation box at a given time $t$. The growth of \alf waves leads to a reduction in the mean free path in the ISM region, resulting in modulations in CR energy flux and densities. After a short transient from $\sim 20-40 \times 10^3 \Omega_c^{-1}$, the distribution adjusts to form an approximately spatially-constant flux that declines over time. In this steady state configuration, a negative (positive) energy density gradient forms in the ISM (cloud), consistent with fluid theory.}
\label{fig:space_time_fiducial}
\end{figure*}

We can gain further insight into the formation of the gradients by studying the space-time evolution of the CR moments (Figure~\ref{fig:space_time_fiducial}). Before $t$~=~4$\times 10^4$~$\Omega_c^{-1}$, the simulations exhibit transient behavior, with fluctuations in the energy density and flux on scales of the cloud. Fluctuations in the energy density are asymmetric about the ISM-cloud interface. Furthermore, the cosmic ray pressure gradient develops slightly earlier in the ISM region, indicating that the cloud gradient is a response to the pressure gradient in the diffusive transport region. 

For a brief interval around $t$ = 3$\times 10^4$ $\Omega_c^{-1}$, a peak in the number/ energy density forms on the upstream side of the cloud. 
This peak forms immediately after the wave amplitudes exceed ${\cal E}_A/\rho_0 v_A^2 \sim 10^{-4}$, i.e. when the scattering rate becomes shorter than the current simulation time. Thus, this peak is a transient effect which forms just as particles begin scattering back into the cloud. The peak is soon erased as forward-propagating particles crossing the ISM are scattered away from the leading interface, unable to replenish the deficit left by particles scattered forward into the cloud.

\subsection{Origins of the Gradient: Kinetic Perspective}

The advantage of treating CRs as kinetic particles in the PIC method is that we are able to obtain the full distribution function at any given time or location in the simulation. We choose to output slices in the distribution function in position bins averaged over a width of 1.25$\times$10$^4$ $v_A \Omega_c^{-1}$, corresponding to 16 position bins in the \texttt{Fiducial} and \texttt{NCR2} simulations and 40 bins in the \texttt{Long\_Cloud} and \texttt{Long\_ISM} runs. 

\begin{figure*} [!htb]
\hbox{
\psfig{figure=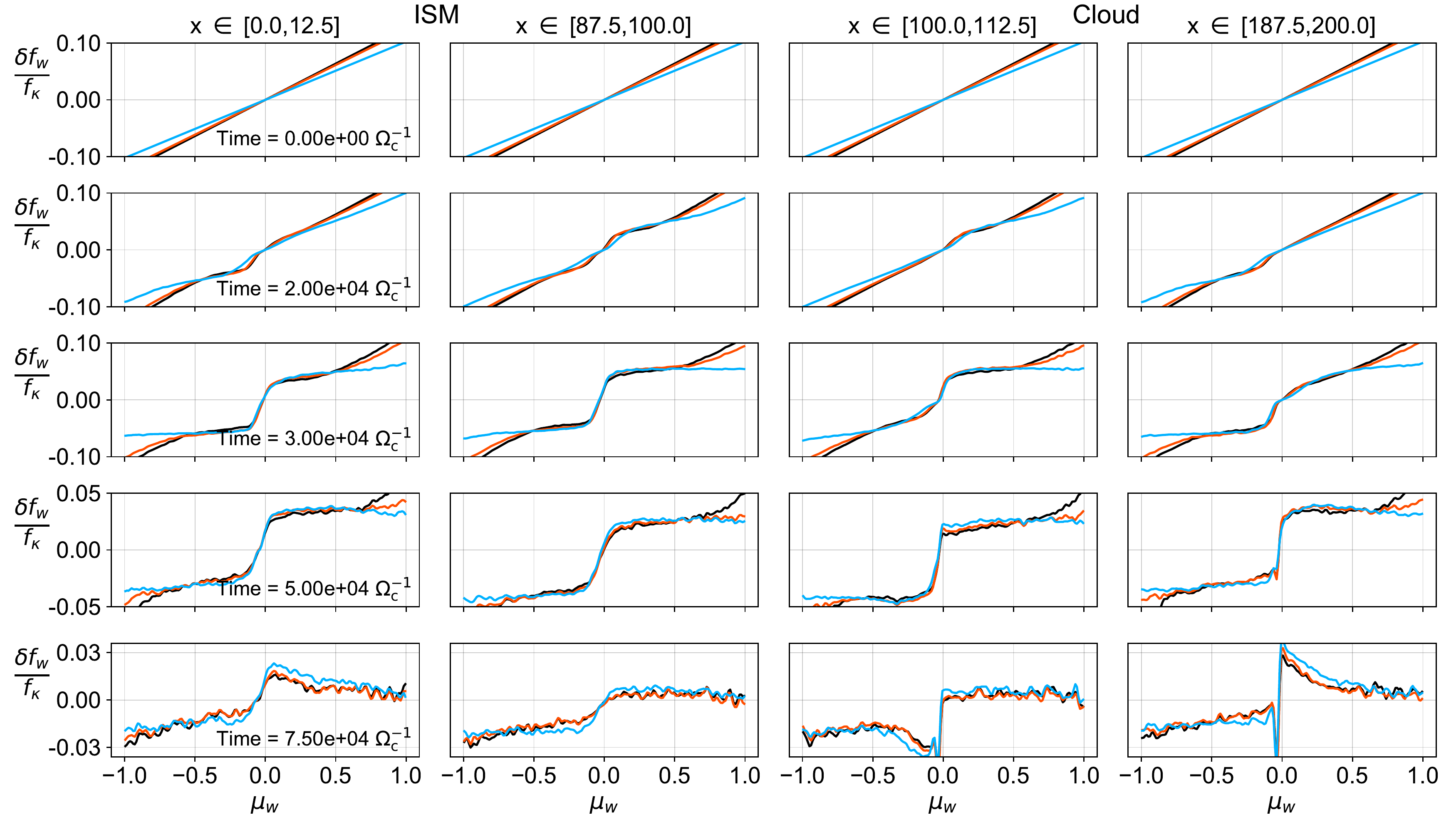,width=1.0\textwidth}
}
\caption{Slices of the wave frame perturbed distribution function $\delta f_w = f_w - f_{\kappa}$ at four locations near the ISM-cloud interfaces. Three momentum bins are shown, corresponding to the peak of the number density (light blue), energy density (orange), and energy flux (black) integrals, at $\log{(p_w/p_0)}$ = 0.19, 0.45, and 0.55 respectively. As time increases (top to bottom), the wave frame distribution moves from an anisotropic drifting state toward an isotropic, $\mu$-independent state except near $\mu=0$. }
\label{fig:distribution_evolution}
\end{figure*}

Figure~\ref{fig:distribution_evolution} shows the space-time evolution of momentum bins of the wave-frame distribution function. We choose to study three momentum bins, corresponding to the peak momenta of the $n_{\mathrm{CR},w}$, ${\cal E}_{\mathrm{CR},w}$, and $F_{\mathrm{CR},w}$ integrals, i.e. $\log{(p_w/p_0)}$ = 0.19, 0.45, and 0.55 respectively. For ease of comparison, we study the quantity $\delta f_w/f_{\kappa}$, where $\delta f_w = f_w - f_{\kappa}$. This perturbation is slowly erased as the distribution function relaxes toward an isotropic, approximately $\kappa$ distribution in the wave frame. 

Initially, $\delta f_w/f_{\kappa}$ increases linearly with $\mu_w$, antisymmetric about $\mu_w=0$, i.e. the distribution has a net flux (Row~1). As waves grow and scattering begins (Row~2), the distribution evolves differently at the leading and trailing interfaces, respectively downstream and upstream from the ISM. At the leading interface (central two columns), particles with positive pitch angle arrived by crossing through the ISM. These forward-propagating particles were scattered toward smaller pitch angle, leading to the development of a shelf-like structure near $\mu_w = 0.2$ at the expense of particles near $\mu_w$~=~1 (Row~3). A similar phenomena occurs at the trailing interface. Backward-propagating particles are scattered toward more negative pitch angle, increasing the deficit near $\mu_w = -0.2$ and reducing the deficit near $\mu_w = -1$. The deficit near $\mu_w = -0.2$ is only filled in when forward-propagating particles can cross the 90$^{\circ}$ barrier. 

As time progresses, a common story emerges in the ISM. We focus on forward-propagating particles since they are the dominant population. Particles entering from the trailing cloud-ISM interface ($x$~=~0) are scattered as they cross the ISM. Upon arriving at the leading cloud interface ($x=100$ in units $10^4 v_A \Omega_c^{-1}$), they are suddenly able to free stream into the cloud, rapidly vacating the interface region. This sudden transition leads to a net deficit just upstream of the  leading ISM-cloud interface, and a steep gradient forms near $x=100$ (Row~4). Particles built-up near $\mu_w \gtrsim 0$ in the cloud (Row~5) are scattered over the 90$^{\circ}$ barrier as they cross the ISM, providing fewer particles to replenish the deficit left by the now free-streaming CRs entering the cloud; the gradient is enhanced. These effects are compounded by the continuing decrease in the CR flux, providing fewer and fewer particles to fill in the deficit with time. Thus, we are left with an excess at $x$ = 0 and a deficit at $x=100$. This process is the essence of the spatial diffusion which yields a gradient in the CR number and energy densities. 

\begin{figure*} [!htb]
\hbox{
\psfig{figure=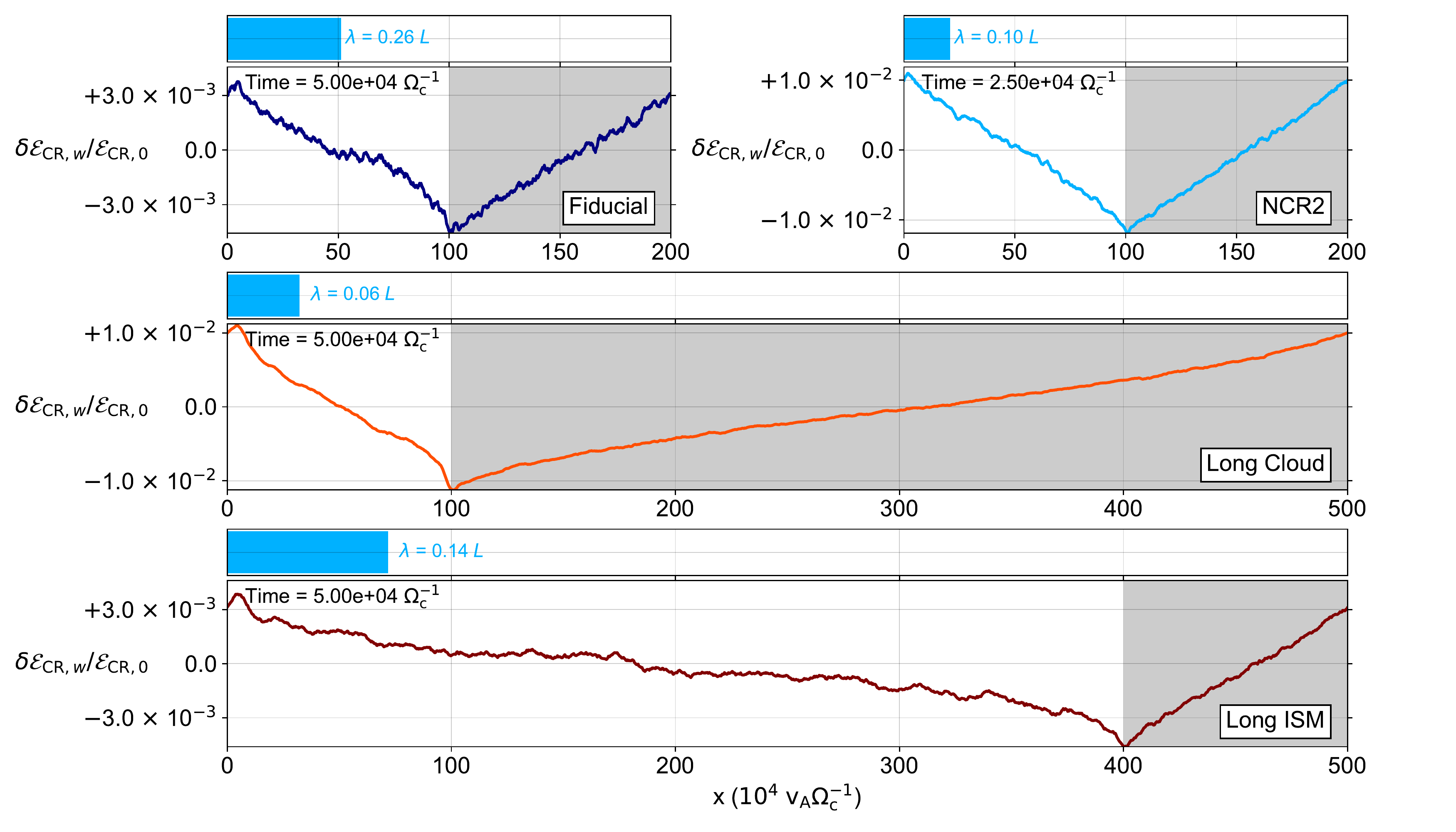,width=1.0\textwidth}
}
\caption{Energy density profiles and mean free paths for all simulations. Times are selected such that the gradient is maximized in the simulation. 
In all simulations, the peak of the energy density perturbation is near the 
upstream side of the ISM (trailing end of cloud)
$x$ = 0 while the minimum appears downstream from the ISM (leading end of cloud). The \texttt{NCR2} and \texttt{Long\_Cloud} gradients in the ISM are the largest, implying that larger wave amplitudes (scattering rates) yield more substantial gradients, consistent with fluid theory.}
\label{fig:gradient_comparison}
\end{figure*}

The positive gradient in the cloud is formed by a different process. We see the emergence of a peak in the distribution function of forward-propagating particles near $\mu_w=0$ as particles cross the cloud. Since low amplitude waves in the cloud cannot scatter particles at any substantial rate, the distributions in the cloud are purely inherited from the ISM. The peak apparent in the rightmost column of Figure~\ref{fig:distribution_evolution} can be explained by considering the crossing time of particles through the cloud. CRs with pitch angles of $\mu_w$ = 0.1 have 10\% of the parallel velocity of those with $\mu_w$ = 1 for fixed momentum. Thus, particles near $\mu_w=0$ entered the cloud at an earlier time in the simulation, when the overall CR flux was higher. CRs near $|\mu_w|$ = 1 entered the cloud recently by comparison, after the CR flux dropped substantially. Thus, we are seeing an overlap of wave-frame distributions at different times, and this overlap effect becomes increasingly clear near the trailing interface ($x$~=~0).

By studying the distribution functions, we find a kinetic explanation for the same behavior predicted by fluid theory. A negative spatial gradient across the ISM forms in response to wave-particle scattering. The positive gradient across the cloud is a response to the drop in CR energy flux, since particles further in the cloud for the same momentum and pitch angle entered the cloud at an earlier time when the overall CR flux was higher. Together, these processes work to create a pressure balance of the CR ``fluid'' across the contact discontinuity between the ISM and cloud.

\subsection{Comparing Spatial Structure among Simulations}

Before using our simulations to compute the wave-particle scattering rates, we turn our attention to the structure and magnitude of the spatial gradient in CR energy density, which provides an insight into the magnitudes of the wave-particle scattering rates.

Figure~\ref{fig:gradient_comparison} displays the energy density spatial structure and mean free paths in the ISM for all simulations at times when the gradient is near maximal. The \texttt{Fiducial} simulation contains the fewest mean free paths across the ISM ($\sim$2), while the \texttt{Long\_ISM} simulation contains nearly 6 mean free paths.

The overall magnitude of the fluctuation in CR energy density is connected to the amplitudes of \alf waves. This is evidenced by the fact that the simulations with the largest amplitude waves (\texttt{NCR2} and \texttt{Long\_Cloud}) have energy density fluctuations more than a factor of 3 larger than the \texttt{Fiducial} run, despite possessing an ISM region of equal size. Similarly, the fluctuations in the \texttt{Fiducial} and \texttt{Long\_ISM} runs are approximately equal, consistent with the similarity in \alf wave energy density between these simulations (Figure~\ref{fig:wave_energy_flux}).

\begin{figure*} [!htb]
\hbox{
\psfig{figure=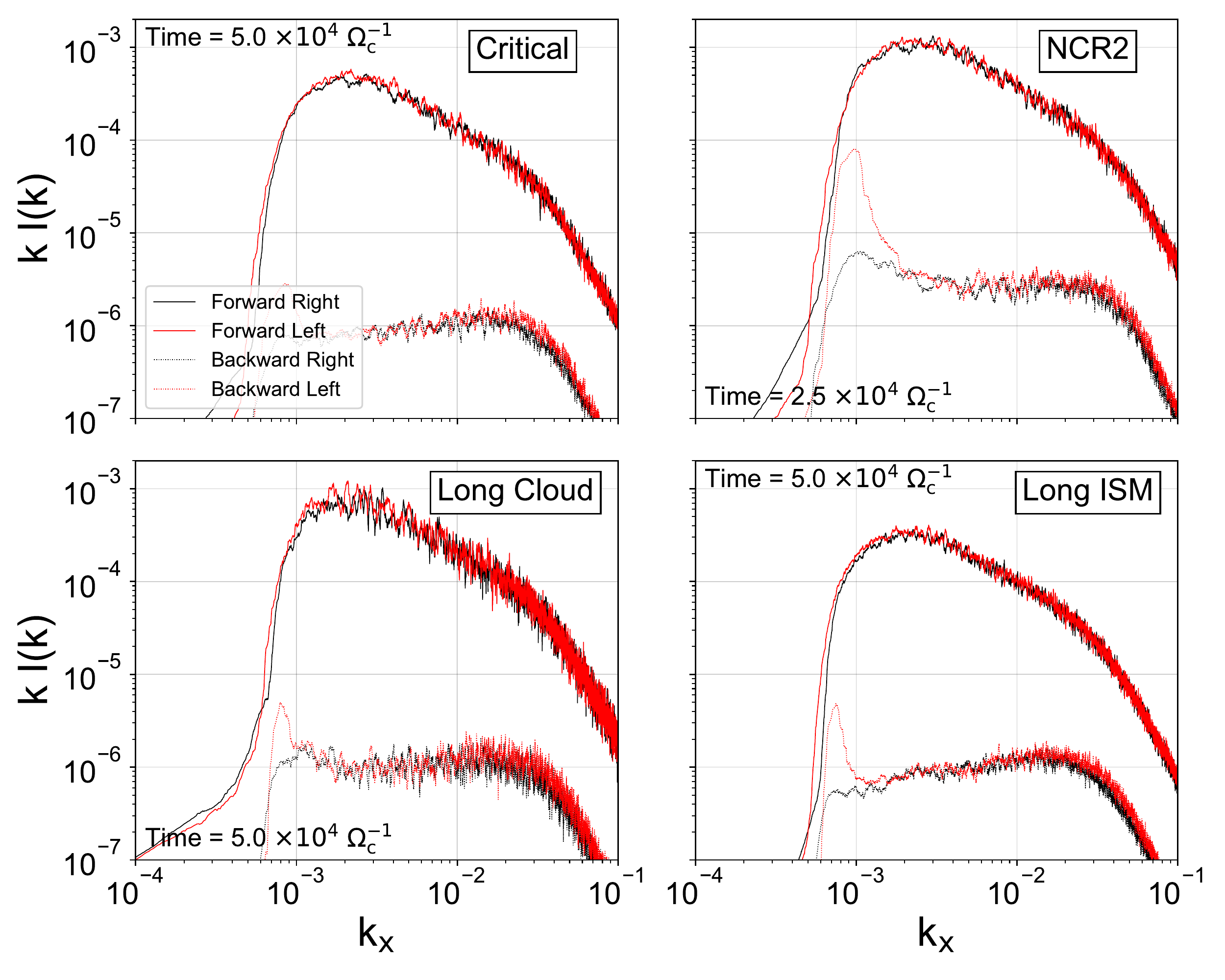,width=1.0\textwidth}
}
\caption{ISM \alf wave power spectra at times when the CR energy density gradient is near maximal. For plotting purposes, these spectra are smoothed on a length scale of 100$k_L$, where $k_L v_A/\Omega_c$ = 2$\pi$/$L$ = $\pi$ and $\frac{2}{5} \pi$ $\times 10^{-6}$ for the \texttt{Critical}/\texttt{NCR2} and \texttt{Long\_Cloud}/\texttt{Long\_ISM} simulations respectively. Note that this smoothing length is 5$\times$ larger than that used for computing the effective quasilinear scattering rate. Forward propagating waves contain the majority of power, shared equally between left and right polarizations. The self-generated wave spectrum from our $\kappa$ distribution of CRs has a slope $I(k) \sim k^{-2}$ at scales just below the peak scale near $k v_A/ \Omega_c$ = $v_A$/$p_0$ $\approx$ 3$\times 10^{-3}$.}
\label{fig:power_spectra}
\end{figure*}

Comparing the steepness of gradients in the ISM and cloud for a given simulation provides insight into the relative importance of each term in the fluid equation (\ref{eq:fluid_equation}).  For the \texttt{Fiducial} and \texttt{NCR2} runs, the gradients must be equal in magnitude in the ISM and cloud, because they are of equal length, $L_{\mathrm{ISM}} = L_{\mathrm{Cloud}}$.  In the cloud region, the energy density gradient must be equal in magnitude (and opposite in sign) to the time derivative in energy flux because the scattering rate is negligible.  Therefore, 
the two terms on the LHS of Equation~\ref{eq:fluid_equation} must be equal in magnitude for these two models.   
Following similar reasoning, one would expect the flux time derivative term to be $\sim 1/4$ of the pressure gradient term for the \texttt{Long\_Cloud} model, and the flux time derivative term to be $4$ times the pressure gradient term for the \texttt{Long\_ISM} model. Thus, modulating the ISM-to-cloud length ratio, $L_{\mathrm{ISM}}/L_{\mathrm{Cloud}}$ explores the relative importance of each term and provides a test of robustness for the fluid theory when we compute the fluid scattering rate in Section~\ref{sec:fluid_scattering}.

\begin{figure*} [!htb]
\hbox{
\psfig{figure=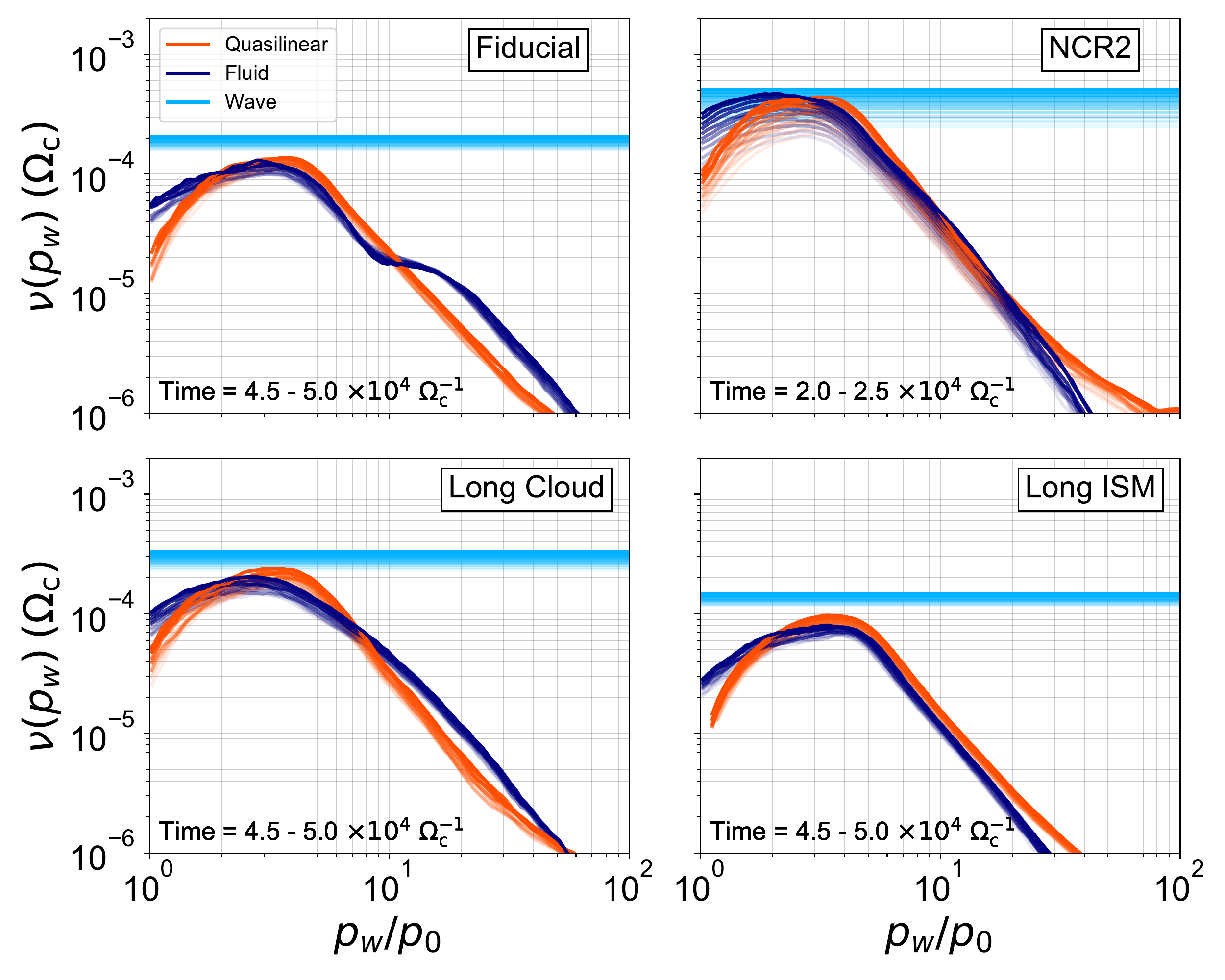,width=1.0\textwidth}
}
\caption{Effective scattering rates computed from quasilinear theory (Equation~\ref{eq:effective_scattering1}; orange) and fluid theory (Equation~\ref{eq:fluid_scattering}; dark blue) as a function of momentum for all simulations. The naive estimate based on the average \alf wave energy density (Equation~\ref{eq:naive_scattering}; light blue) is over-plotted for comparison. Curves are smoothed in momentum for plotting purposes. We show the time evolution of the scattering rates by varying the transparency of the curves with time; the most transparent lines are from the earliest measurement time and the darkest lines are from the latest measurement time within the range displayed.} 
\label{fig:scattering_comparison}
\end{figure*}

So far we have focused on the energy density structure in the ISM, which always has a negative gradient (as expected for a diffusive region with positive flux); however, a different process operates in the cloud. For all of our models, the energy density shows a consistent positive gradient across the cloud. This increase is a consequence of the continual decrease in energy flux with time in all the simulations. As the CRs approach isotropy, the rate of change of the  flux decreases (Figure~\ref{fig:wave_energy_flux}b), and the energy density gradient in the cloud is slowly erased (Figure~\ref{fig:space_time_fiducial}). In the present simulations, the temporal decrease in flux is a consequence of periodic boundary conditions. In the real ISM, we would instead expect an approximate steady state for the flux to be reached. A state with a temporally constant flux and negligible scattering within a dense, neutral cloud (due to strong local damping) would not have a spatial gradient in the energy density within the cloud. Rather, a downward energy density ``ramp'' in the diffusive ISM region would be followed by an energy plateau within the cloud. The emergence of the CR density gradients in the cloud, while entirely consistent with fluid theory, is an artifact of the periodic boundary condition in the present simulations. 

\section{Scattering Rate Comparison} \label{sec:scattering_rates}

In this section, we present a numerical verification of the moment equation which underpins CR hydrodynamics. We compare the wave-particle scattering rates as predicted by quasilinear and Fokker-Planck theory to show that these methods agree well with CR hydrodynamics when the mean free path to scattering is sufficiently short. Since the scattering rate encodes spatial diffusion, this section serves as first-principles confirmation of the quasilinear calculation for the parallel CR diffusion coefficient in fluid theory. Deviations from the quasilinear prediction point to nonlinear effects, particularly near the $\mu=0$ barrier (see Sections~\ref{sec:CR_hydrodynamics} and~\ref{sec:nonlinear_scattering}).

\subsection{Effective Quasilinear Scattering Rate}

We compute both the effective quasilinear and fluid scattering rates using the full distribution function measured in the wave frame $f_w$. Equation~\ref{eq:effective_scattering1} provides the procedure for computing the effective scattering rate from quasilinear theory. Following \citetalias{Bai2019}, we decompose our \alf waves into left/right-handed and forward/backward-propagating modes, 4 power spectra in all. We compute these power spectra based on the waves in the ISM only (zero-padding the cloud) and normalize the spectra according to Equation~\ref{eq:power_spectrum}. Thus, the integrated power spectra together equal the quantity $(\delta B_{\bot}/B_0)^2$ averaged over the ISM. Power spectra are then smoothed over a scale of 20$k_L$, where $k_L$~=~$2 \pi/L$ is the wavenumber corresponding to the box size $L$. We find that less smoothing of the power spectra in $k$-space yields too much noise in the scattering rate which causes the integral in Equation~\ref{eq:effective_scattering1} to be poorly behaved. The chosen smoothing does not change the shape of the spectrum. Smoothed \alf wave spectra for all simulations are shown in Figure~\ref{fig:power_spectra}. This procedure allows us to compute the quasilinear scattering rate through Equation~\ref{eq:quasilinear_scattering}.

To obtain the \textit{effective} scattering rate from $\nu_{QL}$ and $f_w$, we smooth $f_w$ and average the distribution over the ISM, omitting the spatial bins nearest the ISM-cloud boundaries. This omission eliminates particles trapped near $\mu=0$ (see Figure~\ref{fig:distribution_evolution} bottom right panels) which can bias the $\mu$-derivative near $\mu=0$; however, including these bins simply increases the noise in the scattering rate integral. We compute the $\mu$-derivative of $f_w$ using a centered finite difference and perform the integrals in Equation~\ref{eq:effective_scattering1} over pitch angle and gyrophase only, leaving scattering rates as a function of momentum. The same procedure holds for the energy flux in the denominator of $\nu_{\mathrm{eff}}$ according to Equation~\ref{eq:effective_scattering1}. The results are shown as the orange curves in Figure~\ref{fig:scattering_comparison}.

\subsection{Fluid Scattering Rate} \label{sec:fluid_scattering}

 The fluid scattering rate is computed from the moments of $f_w$. First, the energy-weighted moments of the distribution function are calculated according to Equations~\ref{eq:energy_density} and~\ref{eq:energy_flux}. The integrations are performed over pitch angle and gyrophase alone such that the resulting expressions represent infinitesimal moments as a function of momentum. These moments are then inserted into the fluid equation, Equation~\ref{eq:fluid_equation}. The time derivative of the energy flux is computed by measuring the flux at 200 output times over the simulation time of 10$^5$~$\Omega_c^{-1}$, corresponding to an interval of 500 $\Omega_c^{-1}$ between outputs. Then, a second order accurate centered finite difference is used to compute the time derivative. The energy density gradient is computed by performing a linear fit to the energy density as a function of position $x$ in the ISM, comprising 8 spatial bins for the wave spectrum output in the \texttt{Fiducial} and \texttt{NCR2} simulations, and 20 spatial bins in the \texttt{Long\_Cloud} and \texttt{Long\_ISM} runs. The slope of this fit is the energy density gradient. We handle the energy flux term on the RHS of Equation~\ref{eq:fluid_equation} by averaging the energy flux over the ISM. Rearranging Equation~\ref{eq:fluid_equation} then yields the following form for the fluid scattering rate,
 \begin{equation} \label{eq:fluid_scattering}
     \nu_{\mathrm{fluid}} (p_w) = - \frac{\frac{\partial F_{\mathrm{CR}} }{\partial t} (p_w) + \frac{v_w^2}{3} \frac{\partial {\cal E}_{\mathrm{CR}}}{\partial x} (p_w)}{F_{\mathrm{CR}} (p_w)},
 \end{equation}
 where $\nu_{\mathrm{eff}}$ has been replaced by $\nu_{\mathrm{fluid}}$ to eliminate ambiguity with the effective quasilinear scattering rate. This fluid scattering rate as a function of momentum for all simulations is shown in Figure~\ref{fig:scattering_comparison}. Note that each term in Equation~\ref{eq:fluid_scattering} can be computed from the CR energy density gradients (Figure~\ref{fig:gradient_comparison}) and CR flux. The time rate-of-change in the flux must be balanced by the energy density gradient in the cloud region, where $\nu_{\mathrm{eff}} \approx 0$. Thus the cloud gradient provides the time derivative term while the ISM gradient provides the position derivative.

In general, we find good agreement between the effective scattering rates computed via fluid and quasilinear theory. The scattering rates do not differ by more than 50\% over more than an order of magnitude in momentum, 1$<p_w/p_0<$20. For much of this range  (1$<p_w/p_0<$10) the deviation is within 25\%.

Except in the case of very strong scattering (simulation \texttt{NCR2}), the fluid and effective quasilinear scattering rates are lower than the naive estimate for the scattering rate (Equation~\ref{eq:naive_scattering}; light blue line in Figure~\ref{fig:scattering_comparison}) by a factor of order unity. The naive estimate, which relies on a strict random walk in pitch angle, is most accurate in the regime of strong scattering when particles are less affected by the $\mu=0$ barrier (see Section~\ref{sec:CR_hydrodynamics}).

Given that the quasilinear scattering rate $\nu_{QL} (p_w,\mu_w)$ and certainly the distribution function suffer from particle discreteness noise while the fluid scattering rate includes nonlinear effects (see Sections~\ref{sec:nonlinear_scattering}), these minor deviations are within the range of accuracy we expect to achieve from our PIC simulations. In addition, we note that the integral in Equation~\ref{eq:effective_scattering1} is poorly behaved and runs into a ``sign problem'' for lower particle momenta, yielding a negative effective scattering rate---an unphysical result. These scattering rates are omitted in Figure~\ref{fig:scattering_comparison}. The agreement between the scattering rates indicates that, for particles with momenta near the peak of the distribution where the mean free path is shorter relative to the tails of the distribution, the quasilinear prediction is borne out by the fluid behavior of the CRs.

\begin{figure*} [!htb]
\hbox{
\psfig{figure=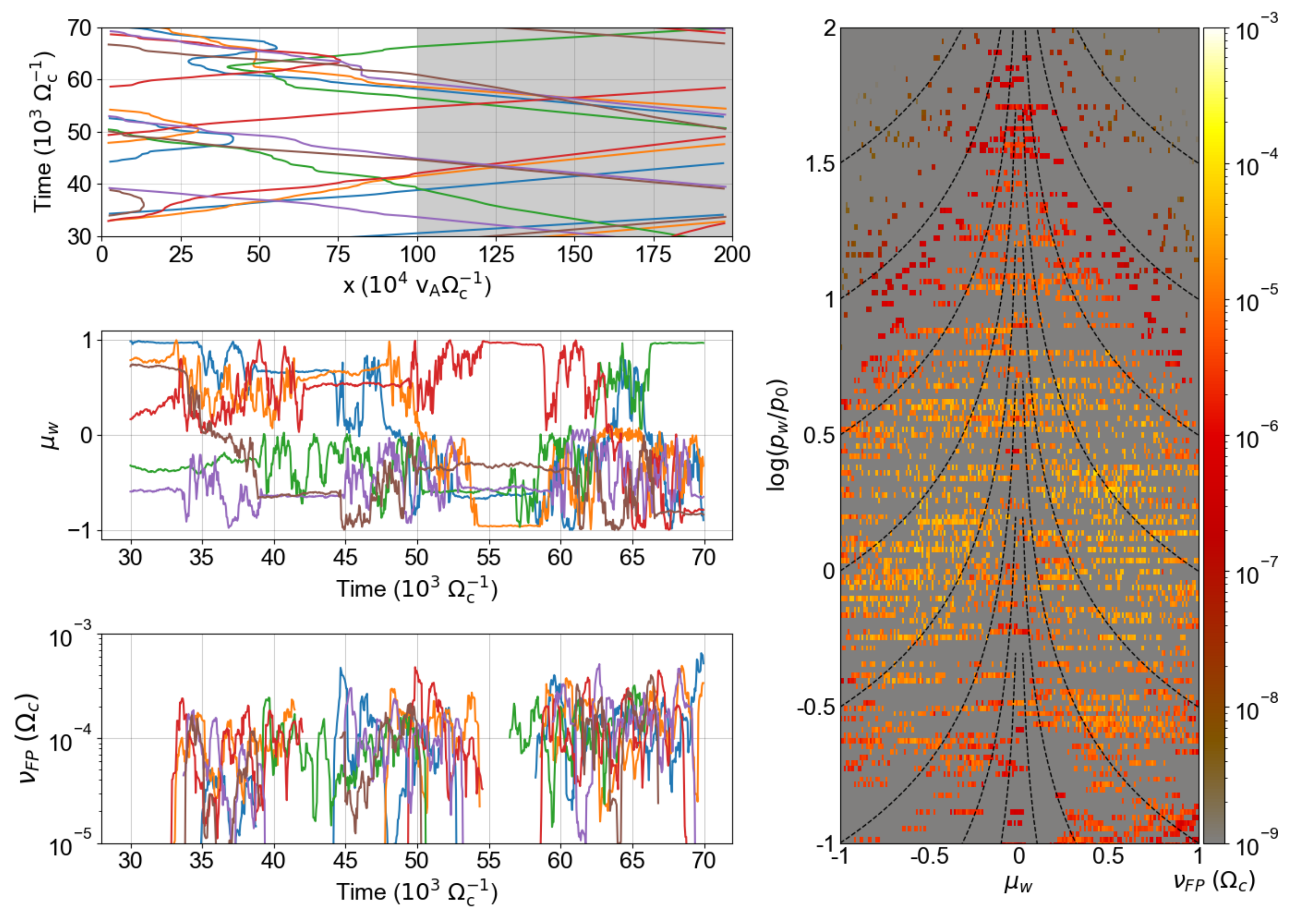,width=1.0\textwidth}
}
\caption{Left: Tracked particle orbits from \texttt{Fiducial} simulation evolving in space (top), pitch angle (middle), and scattering rate (bottom). Right: Fokker-Planck scattering rate at time $t$ = 5$\times$10$^4$ $\Omega_c^{-1}$ as a function of wave-frame momentum $p_w$ and pitch angle $\mu_w$. Note that the top left plot is a space-time plot, where position is on the $x$-axis, consistent with the simulation. Particles are selected such that they appear near the trailing cloud-ISM interface ($x$ = 0) at time $t$ = 5$\times$10$^4$ $\Omega_c^{-1}$. Orbits evolve within the periodic domain, scattering in the ISM region and free streaming through the cloud (grey region). All scattering rates shown are restricted to the ISM only, as cloud scattering is negligible.}
\label{fig:particle_orbits}
\end{figure*}

\subsection{Fokker-Planck Coefficient}
\label{sec:FP_coefficient}

The quasilinear diffusion equation (\ref{eq:quasilinear_diffusion}) is a Fokker-Planck equation with the quasilinear scattering rate weighted by $(1-\mu^2)/2$ acting as the momentum diffusion coefficient $D_{\mu \mu}$. Because of this correspondence, we expect particles to diffuse in pitch angle at a rate given by Equation~\ref{eq:Fokker_Planck_scattering}. In this subsection, we present our procedure for tracking individual particles and computing their scattering rates. This method provides a means of comparing the quasilinear scattering rate (Equation~\ref{eq:quasilinear_scattering}) with the Fokker-Planck rate. In this way, we test the validity of quasilinear theory in a discrete sense, rather than the statistical ensemble captured by fluid theory. 

We track a total of 3200 particles in each simulation. These particles are sampled equally from 200 spatial bins (16 particles per spatial bin). The range of momenta in the drifting (initially isotropic) frame, $-2$~$<\log{(p_d/p_0)}<$~2, is divided into 8 bins. We sample an equal number of particles from each of these 8 bins, drawn from the initial $\kappa$ distribution. Thus, the overall sampled distribution is not a $\kappa$ distribution, but rather 8 distinct momentum ranges within which the particles obey a $\kappa$ distribution (400 particles per momentum bin). In pitch angle, we sample from the initially flat drift frame distribution, tracking forward and backward particles of the same $|\mu_d|$. For instance, if we track particle $i$ with phase space coordinates ($p_{d,i}$,$\mu_{d,i}$,$x_i$), we also track a particle with coordinates ($p_{d,i}$,$-\mu_{d,i}$,$x_i$). This procedure ensures an equal number of forward and backward-propagating particles in the drift frame, which when boosted to the wave frame, yields more particles at positive than negative pitch angle in the wave frame.

Particles' full phase space coordinates (position and momentum) are output every $\Delta t_{FP}$~=~50$\Omega_c^{-1}$, implying that particles with $p \approx p_0$ undergo 50/$\gamma$ $\approx$ 35 gyrations about the magnetic field between outputs. If a particle $i$ is measured at a time $t$, we compute the scattering rate according to Equation~\ref{eq:Fokker_Planck_scattering} as

\begin{equation} \label{eq:discrete_Fokker_Planck}
    \nu_{FP} (t,p_{w,i},\mu_{w,i}) = \frac{\left( \mu_{w,i} (t + \Delta t_{FP}) - \mu_{w,i} (t) \right)^2}{2 \Delta t_{FP}}.
\end{equation}

\indent Figure~\ref{fig:particle_orbits} displays particle positions, pitch angles, and scattering rates as a function of time for a selection of 6 particles in the \texttt{Fiducial} simulation which undergo a 90$^{\circ}$ pitch angle crossing. We select particles such that they arrive near the peak of the CR energy density gradient at $t$~=~5$\times$10$^4$ $\Omega_c^{-1}$ from the bin 0$<\log{(p_w/p_0)}<$0.5. Note that even though these particles are near the peak of the distribution, they undergo very few direction reversals, and diffusion is primarily in pitch angle rather than real space during a single crossing of the ISM.

\begin{figure*} [!htb]
\hbox{
\psfig{figure=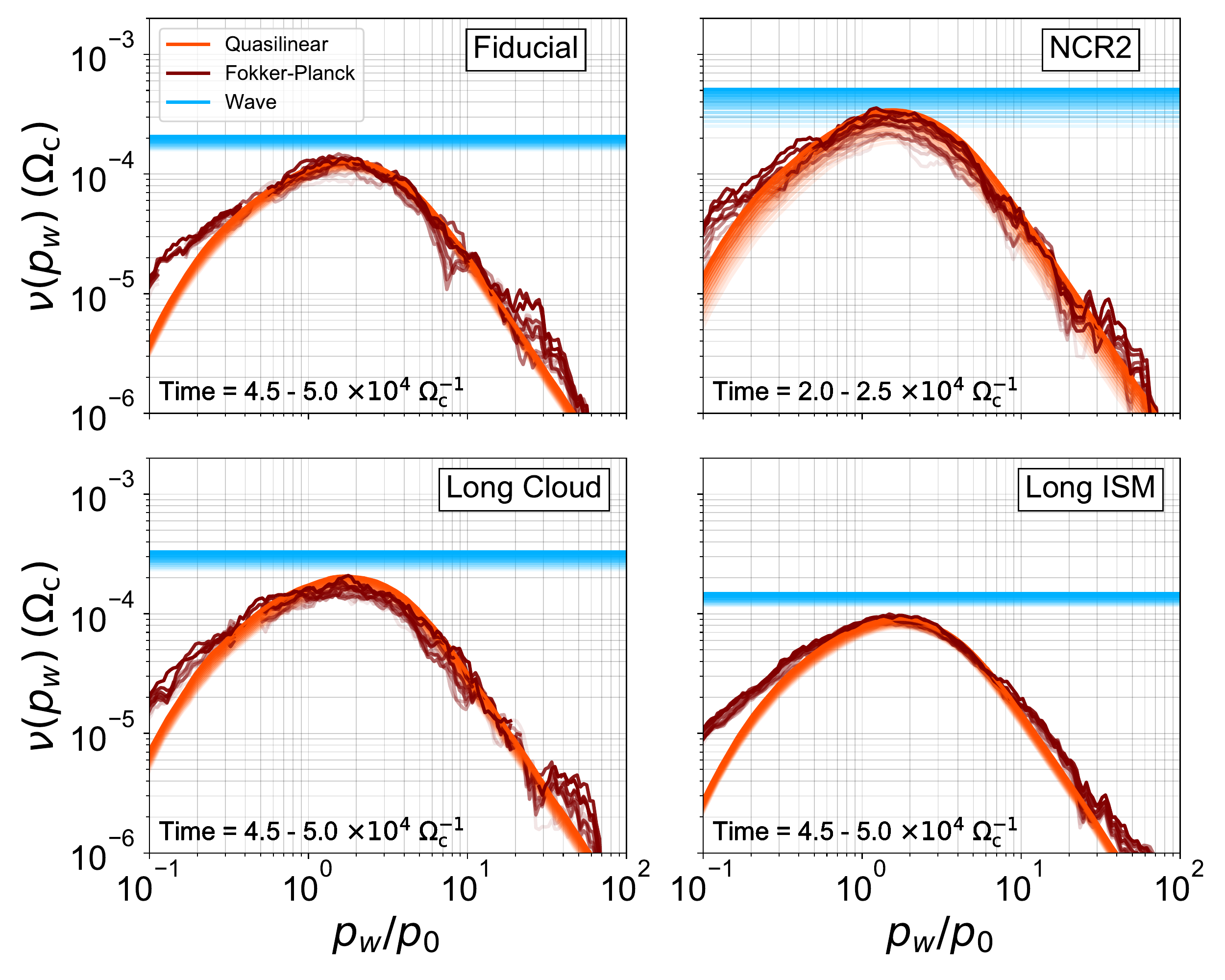,width=1.0\textwidth}
}
\caption{Ensemble averaged scattering rates computed from Fokker-Planck theory (Equations~\ref{eq:discrete_Fokker_Planck} and~\ref{eq:ensemble_average}; dark red) and quasilinear theory (Equations~\ref{eq:quasilinear_scattering} and ~\ref{eq:weighted_average}; orange) as a function of momentum for all simulations. These scattering rates agree remarkably well across nearly 3 orders of magnitude in momentum, with exquisite concordance at the peak of the CR momentum distribution. Similarly, these scattering rates agree well with the naive estimate (Equation~\ref{eq:naive_scattering}; light blue) at this peak.} 
\label{fig:FP_scattering}
\end{figure*}

Scattering rates center around 2--3$\times$10$^{-4}$ $\Omega_c$, although rates can fluctuate up to 10$^{-3}$ $\Omega_c$ for the particles tracked. A better sense of the full scattering rate distribution is shown in the right panel of Figure~\ref{fig:particle_orbits}. Clearly, 3200 particles is a sparse sampling of the full distribution function (this is seen most clearly by juxtaposing this figure with Figure 9 of \citetalias{Bai2019}). Yet, we still see clear patterns emerging, namely an increase in scattering rate near the peak of the distribution around $p_0$ and a decrease in scattering rate with increasing momentum. Unfortunately, the sparseness does not allow us to see constant scattering rates along resonant lines $k_{\mathrm{res}}$ = constant (dotted black lines); however, we see an approximately constant scattering rate as a function of $\mu_w$ for fixed momentum. The exception to this pattern is near $\mu_w=0$, where the scattering rate drops to $\sim$10$^{-6}$~$\Omega_c$.

With the scattering rate distribution shown in the right panel of Figure~\ref{fig:particle_orbits}, we are able to compute a pitch angle-averaged scattering rate which we can compare to that from quasilinear theory (Equation~\ref{eq:quasilinear_scattering}). We define this ensemble average,
\begin{equation} \label{eq:ensemble_average}
    \left< \nu_{FP} \right> (p_w) = \frac{\int \int \nu_{FP} (p_w,\mu_w,x) f_w (p_w,\mu_w,x) \: d \mu_w dx}{\int \int f_w (p_w,\mu_w,x) \: d \mu_w dx}.
\end{equation}
Here, the integral in $\mu_w$ is over the full range of pitch angle, $-$1 to $+$1, and the integral in $x$ is over the ISM region. Note that as was done with the effective quasilinear scattering rate, we omit the spatial bins nearest the cloud-ISM boundaries since particles retain their distribution from the free-streaming cloud in these regions. Because the distribution is sparse, we set all momentum and pitch angle bins in $f_w$ and $\nu_{FP}$ to 0 where no particle is present in that element of phase space. 

Comparing this Fokker-Planck rate to quasilinear theory requires a different weighting of the quasilinear scattering rate $\nu_{QL}$. Transforming to pitch angle coordinates introduces a geometric factor of $(1-\mu_w^2)/2$ to the quasilinear scattering rate in Equation~\ref{eq:quasilinear_diffusion}. Thus, the proper average of the quasilinear scattering rate is given by
\begin{equation} \label{eq:weighted_average}
     \left< \nu_{QL} \right> (p_w) = \frac{\int \left(\frac{1 - \mu_w^2}{2} \right) \nu_{QL} (p_w,\mu_w) f_w (p_w,\mu_w) \: d \mu_w}{\int  f_w (p_w,\mu_w) \: d \mu_w},
\end{equation}
where this expression is obtained from Equation~\ref{eq:FP_QL} and $\nu_{QL}$ is given by Equation~\ref{eq:quasilinear_scattering}. Note that because the quasilinear scattering rate is a function of the \alf wave power spectrum as measured over the entire ISM, the scattering rate is already averaged over $x$. In the above integral, we average $f_w$ over $x$, again omitting spatial bins nearest the cloud-ISM boundaries.

The ensemble averaged Fokker-Planck scattering rate is compared to the suitably weighted quasilinear scattering rate in Figure~\ref{fig:FP_scattering}. Across a wide range of momenta, nearly three orders of magnitude, these scattering rates agree remarkably well, with near perfect agreement at the peak of the CR momentum distribution. In addition, these scattering rates agree well with the naive scattering rate estimate from Equation~\ref{eq:naive_scattering}. Some deviations are present, most notably at momenta below $p_w/p_0$ = 0.5 and above $p_w/p_0$ = 20, where the Fokker-Planck scattering rate is higher than the quasilinear prediction. We address the question of the validity of quasilinear theory as well as the robustness of fluid models of CR transport in the next section.

\section{Discussion} \label{sec:discussion}

By breaking translational symmetry with spatially-dependent ion-neutral damping, our simulations have enabled an exploration of fluid behavior in a collisionless CR population under the sole influence of self-induced wave-particle interactions. This result is one of a growing number of studies in which collisionless wave-particle interactions act to replace the particle-particle collisions underpinning transport in MHD fluids. 
Whether in collisionless shocks \citep{Spitkovsky2008}, kinetic turbulence \citep{Howes2008,Meyrand2019}, or MRI-unstable shear flows \citep{Kunz2016}, wave-particle interactions regulate transport, heating, and fluid-scale structure. Yet, while fluid behavior might emerge, the precise value of transport coefficients may deviate substantially from collisional or weakly collisional predictions \citep{Spitzer1962,Braginskii1965} or take on a functional form ill-suited to fluid models \citep[Arzamasskiy et al. 2021, in prep.;][]{Kunz2014}. 

Using our simple toy model for a sharp boundary between a mostly-neutral cloud and well-ionized plasma, we directly demonstrate spatiotemporal behavior consistent with that expected from the energy flux equation of CR hydrodynamics. The effective scattering rates in the fluid theory are consistent with the fully kinetic quasilinear prediction and supported by studies of individual particle motions. Thus, for the restrictive case of field-parallel CR diffusion subject only to scattering from small-scale, self-generated \alf waves for our chosen parameters, the quasilinear prediction for the diffusion coefficient is likely accurate. In this section, we address the validity of the quasilinear approximation, the role of nonlinear effects in transport, and how fluid theory remains so robust in our simple system. 

\subsection{Validity of Quasilinear Theory}
\label{sec:quasilinear_theory}

Quasilinear theory, when applied to the gyroresonant CRSI, treats the growth of \alf waves based on a static, drifting distribution function $f_0 (\boldsymbol{p})$, i.e. through \textit{linear} theory. The growing waves induce scattering among the particles, modifying the distribution function. In general, the evolution of the distribution function is a complex, fully nonlinear problem computed numerically in our simulations; however, if particle and wave dynamics occur on disparate timescales, we can use a scale separation technique to evolve the distribution function. \textit{Quasi}-linear theory assumes that the overall distribution function evolves slowly compared to the dynamical timescale for the waves. In essence, it posits a separation of timescales, $k v_A \gg |\partial\ln f_0/\partial t|$. This relation is usually well satisfied for small wave amplitudes, as the ratio between the two scales is $(v_A/c)(\delta B/B_0)^{-2}\gg1$ given our choice of parameters, with $\delta B/B_0\sim10^{-2}$, $c$~=~300~$v_A$.

A critical assumption underlying quasilinear theory is the
``random phase approximation'', i.e. fields are delta-correlated in $k$-space and time \citep{Kulsrud2005}. When this assumption is satisfied, particles experience Gaussian white noise forcing and freely diffuse in momentum space as a central limit effect. Quasilinear theory applies as long as particles encounter sufficiently many uncorrelated wave packets such that they undergo chaotic diffusion in momentum space (see \cite{Besse2011} for further discussion). In our simulations, this condition is met when the mean free path is short relative to the ISM scale. As long as multiple mean free paths fit within the ISM, particles experience multiple scattering events while traversing the ISM and chaotic diffusion ensues. Nonlinear wave-wave interactions may distort the waves and lead to complications which violate the random phase approximation (see Section~\ref{sec:nonlinear_scattering}).  

Away from the peak of the CR momentum distribution, the mean free path to scattering is longer than the ISM scale. The quasilinear approximation is inapplicable for these particles, a fact best exemplified in the flattened structure in the fluid scattering rate near $p_w/p_0$~=~10 in the \texttt{Fiducial} simulation (Figure~\ref{fig:scattering_comparison}). This time-dependent feature moves toward higher momentum as the simulation is run longer, indicating that the structure is a consequence of high momentum particles lacking the time to respond to wave growth through many encounters with waves. Similarly, the deviations at low momentum between the fluid and quasilinear rates in Figure~\ref{fig:scattering_comparison} as well as the Fokker-Planck and quasilinear rates in Figure~\ref{fig:FP_scattering}
point toward sources of scattering not captured by a quasilinear treatment.

Despite these deviations for particles away from the distribution peak, (1) our tracked particles undergo chaotic diffusion in pitch angle, (2) the ensemble averaged Fokker-Planck and weighted quasilinear scattering rates agree well near the peak, and (3) our \alf wave amplitudes remain small. Thus, we argue that quasilinear theory is a valid approximation for the growth and saturation of the gyroresonant CRSI in our simulations for CRs near the peak of the distribution.

\subsection{Implications for CR Hydrodynamics}
\label{sec:CR_hydrodynamics}

Transport in CR fluid theory is encoded in the spatial diffusion coefficient, the inverse of which (times $c^2$) corresponds to the fluid scattering rate that we measure. For CR fluid theory to accurately describe the system's dynamics, this diffusion coefficient should incorporate both quasilinear and nonlinear scattering mechanisms. The agreement between our quasilinear predictions and fluid scattering rates requires efficiently crossing the 90$^{\circ}$ pitch angle ($\mu=0$) barrier.

Overcoming this barrier goes beyond quasilinear theory and requires nonlinear effects (see Section~\ref{sec:nonlinear_scattering}). Parameters chosen in our simulations are far from realistic. In particular, the ratio of $n_{\mathrm{CR,0}}/n_i \sim 10^{-4}$ is highly exaggerated compared to the ratio of CR to thermal particle density in the Galaxy, $\sim10^{-9}$ (although in low-ionization regions $n_{\mathrm{CR}}/n_i$ is much higher). While this exaggeration is necessary to make our simulations computationally feasible, the large wave amplitudes unrealistically enhance nonlinear effects.
A more quantitative understanding of the potential dependence of the scattering rate on wave amplitude is needed to confirm that the quasilinear rate is applicable in CR fluid treatments for realistic ISM environments.

In fact, we can already identify deviations of the effective scattering rate (Equation~\ref{eq:effective_scattering1}, based on the quasilinear theory) from the fluid scattering rate (Equation~\ref{eq:fluid_scattering}, which incorporates nonlinear effects) in Figure~\ref{fig:scattering_comparison}, for CR momenta $p\lesssim p_0$. We have noted that calculation of the effective scattering rate from Equation~\ref{eq:effective_scattering1} suffers from numerical noise, but the trend is already evident. This effective scattering rate is weighted by a gradient in the pitch angle distribution, which is maximized near $\mu=0$ (Figure~\ref{fig:distribution_evolution}) where quasilinear scattering rates vanish. 

Since this effective rate would be formally equal to the fluid scattering rate if quasilinear theory fully described wave-particle interactions, deviations between the effective and fluid rates point to the key role nonlinear effects play in determining the total fluid scattering rate, and subsequently, diffusion. We can conclude that nonlinear effects start to dominate the overall fluid scattering rate for particles with $p\lesssim p_0$ in our simulations. 
Even though realistic wave amplitudes would be lower than those in our simulations, nonlinear effects may also be important in 
realistic ISM environments, as has already been suggested from some recent simulations of galaxy formation (e.g., \citealp{Hopkins2020_Transport}).

We also comment that when studying the Fokker-Planck scattering rate, weighting in pitch angle is uniform; the role of nonlinear effects in overcoming $\mu=0$ is not as significantly manifested. Consequently, there is better agreement between the quasilinear and Fokker-Planck rates in Figure~\ref{fig:FP_scattering}. Despite this agreement, the deviation between the two rates becomes more significant for particles with $p\lesssim0.2p_0$ (see Section~\ref{sec:nonlinear_scattering}), indicating that nonlinear effects dominate over a wide range of pitch angles.

Further research is necessary to elucidate the significance and nature of these nonlinear effects in the hope of developing transport equations faithful to all sources of wave-particle scattering. A first step can come from MHD-PIC simulations which yield numerical estimates of spatial diffusion; however, analytic or semi-analytic techniques may be necessary to extend numerical insight into the regimes relevant for the Galaxy.

\subsection{Nonlinear Sources of Scattering} \label{sec:nonlinear_scattering}

We know some nonlinearity must be present for particles to overcome the $\mu=0$ barrier. Here, we discuss some of the most widely considered mechanisms that may contribute to the enhancement of scattering over the quasilinear prediction in our simulations.

Our wave spectrum is determined by the initial CR distribution, not a scale-by-scale transfer of energy as would be expected within an MHD turbulent cascade. While in general, wave-wave interactions are weak at low amplitude, we do observe, as reported in Plotnikov et al. (2021, in prep.), that the spectrum evolves to include high-$k$ modes, eventually achieving a power-law spectrum in intensity. The origin of this cascade is yet to be understood; however, the effect is present even for wave amplitudes which remain well within the linear regime.

One important consequence of this spectral evolution is generation of abrupt features analogous to rotational discontinuities in transverse magnetic fields. Such features, reported in Plotnikov et al. (2021, in prep.) and present in \citetalias{Bai2019} as well as this work, might be a consequence of nonlinear wave steepening into rotational discontinuities \citep{Cohen1974}. Particles encountering such abrupt features effectively see a sudden change of field direction and hence a sudden change in $\mu$ relative to the perturbed field. This scattering mechanism is generally insignificant given our small wave amplitudes, but it can become significant when $\mu$ is close to zero. A reflection can in principle be achieved by $|\mu|\lesssim\delta B/B_0$. This is identified in \citetalias{Bai2019} as the dominant mechanism for overcoming the $\mu=0$ barrier, and is likely also the mechanism responsible in our simulations.

More generally, wave-wave interactions induced by cascades couple modes in $k$-space, yielding correlated spectral and temporal structure in the waves, relevant on the small spatial (high-$k$) scales of low-momentum gyroresonant particles. These correlations break the assumption of delta-correlated fields (the random-phase approximation) underlying quasilinear theory \citep{Kulsrud2005}, and particles no longer experience Gaussian white noise forcing. The distribution function and wave spectrum co-evolve on similar timescales, and wave-particle interactions must be computed by integrating along perturbed particle orbits rather than the zero-order trajectories used in linear and quasilinear theory. These corrections lead to a broadening of gyroresonances, which can alleviate particles of the strict resonance condition restricting passage over $\mu=0$ \citep{Dupree1966,Weinstock1969,Volk1973,Achterberg1981}. Since diffusion coefficients are time integrals over correlation functions \citep{Shalchi2009}, nonlinear effects such as wave-wave interactions and spatially-localized rotational discontinuities modify the scattering rates measured in Sections~\ref{sec:fluid_scattering} and~\ref{sec:FP_coefficient} away from the quasilinear prediction. 

As is pointed out by \cite{Holcomb2019}, the singularity in the resonant wavenumber in Equation~\ref{eq:resonant_wavenumber} is strictly artificial and can be removed by relaxing the magnetostatic approximation $(\omega/k \sim 0)$. The significance of this modification relies on the presence of sufficient power in broadband backward propagating \alf waves \citep{Schlickeiser1989}. Our wave power spectra in Figure~\ref{fig:power_spectra} indicate that forward propagating modes dominate backward propagating modes by nearly three orders of magnitude, which is a natural consequence of the CRSI. Therefore, this effect alone does not assure our particles' passage through $\mu=0$.

Mirror scattering may play some role in crossing $\mu=0$ \citep{Felice2001, Holcomb2019}; however, this mechanism is likely subdominant in our one-dimensional simulations. Gradients in the magnetic field can form magnetic mirrors which adiabatically scatter particles via the mirror force. This effect is non-resonant and does not break conservation of magnetic moment, in contrast to the aforementioned rotational discontinuities. As variation in the field strength in our simulations is due only to wave motions, we never form large spatial gradients in the magnetic field and thus never generate significant mirror-like structures. Mirror scattering would likely gain greater significance in multiple dimensions and requires future studies of the CRSI which go beyond 1D. In addition, mirror structure may naturally be present in background MHD turbulence, associated with transit time damping from fast magnetosonic modes (e.g. \cite{Schlickeiser1998}, \cite{Yan2004}). 

\subsection{Future Directions}

While we have extended the \citetalias{Bai2019} MHD-PIC simulations to a more realistic environment, we remain far from the conditions relevant to the real ISM. Perhaps most crucially, our problem was studied with periodic boundary conditions. Thus, rather than achieve a steady state energy flux of CRs, we were forced to address a time-dependent problem in which the energy flux is continuously decreasing, leading to the formation of a non-physical positive spatial gradient in the cloud region. While this structure is consistent with CR hydrodynamics, it would not develop for a (quasi) steady state in which wave growth is balanced by damping.

Modifying the boundary condition to a constant flux of CRs entering the simulation domain is a key next step towards greater realism for the problem we have studied. This boundary condition would establish a constant CR energy density gradient and therefore a constant scattering rate for the ISM region. If waves are strongly damped within the cloud, we would expect constant energy density in this region. Such a system could provide a laboratory for measuring the spatial diffusion coefficient as a function of $n_{\mathrm{CR,0}}/n_i$, CR pressure gradient, $\nu_{\mathrm{IN}}/\nu_{\mathrm{crit}}$, and $L_{\mathrm{ISM}}/L_{\mathrm{Cloud}}$.

Even with this modification, the problem of CR transport in and around GMCs remains a challenge. Three dimensional turbulent magnetic fields, CR energy losses due to H$_2$ impact ionization  \citep{McCall1998}, and hadronic losses from CR impacts which produce gamma rays through pion decay in the GeV \citep{Yang2014,Tibaldo2015} and TeV \citep{Aharonian2006,Abramowski2016,Abdalla2018} bands may all contribute to controlling CR transport, energy densities, and ionization rates in GMCs. Thus, a full treatment of this problem requires not only the plasma physics of CR transport, but GMC chemistry and CR energy loss mechanisms as well. 

In the absence of three dimensional effects, additional externally-driven MHD turbulence, or CR energy losses,
and for numerically expedient but unrealistic parameter regimes, our work remains but a first step in understanding CR transport in the multiphase ISM. Yet, by taking a first-principles approach for CR diffusion and comparing to fluid treatments, our work opens a path toward future studies of CR transport coefficients in realistic environments. In this way, studies of CRs in systems from GMCs to galaxy clusters can explore astrophysical macroscales while remaining firmly grounded in the plasma physics of the CR microscales.

\section{Conclusion} \label{sec:conclusion}

We have presented the first self-consistent kinetic simulations of CR transport across an inhomogeneous domain, modeling an embedded neutral cloud within the ionized ISM. By breaking translational symmetry through ion-neutral damping of \alf waves in the cloud region, our simulations enable us to see aspects of fluid behavior in the spatial structure of the CR distribution function. In particular, we show that the simulation results are consistent with the predictions of CR hydrodynamics, in which an energy density gradient and time-dependent energy flux work together to balance wave-particle scattering throughout the ISM. In the ISM region, the gradient in the energy density is in the opposite  direction to the net CR flux. We can understand the ISM energy gradient  as the consequence of diffusive propagation imposed by wave-particle scattering.

In the cloud region, where there is negligible scattering, the gradient in the energy density is in the same direction as the net CR flux, since the decrease in time of the flux must be directly balanced by a spatial gradient of the pressure. This behavior is a consequence of our periodic boundary conditions which ensure that energy flux is a constantly decreasing function of time. The cloud contains a superposition of propagating CR distributions, unchanged after they first entered the cloud. 

Structure in the ISM energy density allows us to compute a wave-particle scattering rate based on the spatio-temporal evolution of the CR moments---a fluid approach. We compare this rate to the quasilinear prediction and Fokker-Planck theory based on particle trajectories. Suitably weighted, all of these scattering rates agree near the peak of the CR distribution where multiple CR mean free paths fit within the ISM region.

The agreement we find among scattering rates serves as a first-principles verification of CR hydrodynamics. A diffusion coefficient computed from quasilinear theory is an accurate description of field-parallel transport due to self-generated wave-particle pitch angle scattering. We note that for the parameters studied in this work, nonlinear effects are exaggerated, which enables crossing of the $\mu=0$ barrier at affordable spatial resolution.  

For more realistic environments with lower wave amplitudes, broadening of gyroresonances would be reduced, but other nonlinear wave-particle interactions, both resonant and non-resonant, may become important. 
Despite these uncertainties, the evidence from our work suggests that CR hydrodynamics is a valid model for CR transport in the ionized ISM, with wave-particle scattering naturally leading to fluid behavior of the collisionless distribution of CRs. Our work thus opens a
pathway toward first-principles calibration of CR fluid transport coefficients in the multiphase ISM.

\acknowledgments

We are grateful to Illya Plotnikov for his contributions to numerical tools for this project. CJB is thankful to Ellen Zweibel, Christoph Pfrommer, Matt Kunz, Anatoly Spitkovsky, Cole Holcomb, Bruce Draine, Phil Hopkins, and Russell Kulsrud for advice and encouragement.  CJB is supported by the NSF Graduate Research Fellowship and the Churchill Foundation of the United States.  XNB acknowledges support by NSFC grant 11873033.  The work of ECO  was supported by grant 510940 from the Simons Foundation. This work began at the \textit{Multiscale Phenomena in Plasma Astrophysics} program at KITP in Santa Barbara, CA. Computational resources for our simulations were provided by the Princeton Institute for Computational Science and Engineering
(PICSciE) and the Office of Information Technology’s High Performance Computing
Center at Princeton University.

\software{{\tt Athena} \citep{Stone2008,Stone2009}, {\tt yt} \citep{Turk2011}}

\vspace{5mm}
\bibliography{CR_Cloud}{}
\bibliographystyle{aasjournal}

\appendix

\section{Selecting the Box Size} \label{sec:box_size}

\begin{figure*}
\hbox{
\psfig{figure=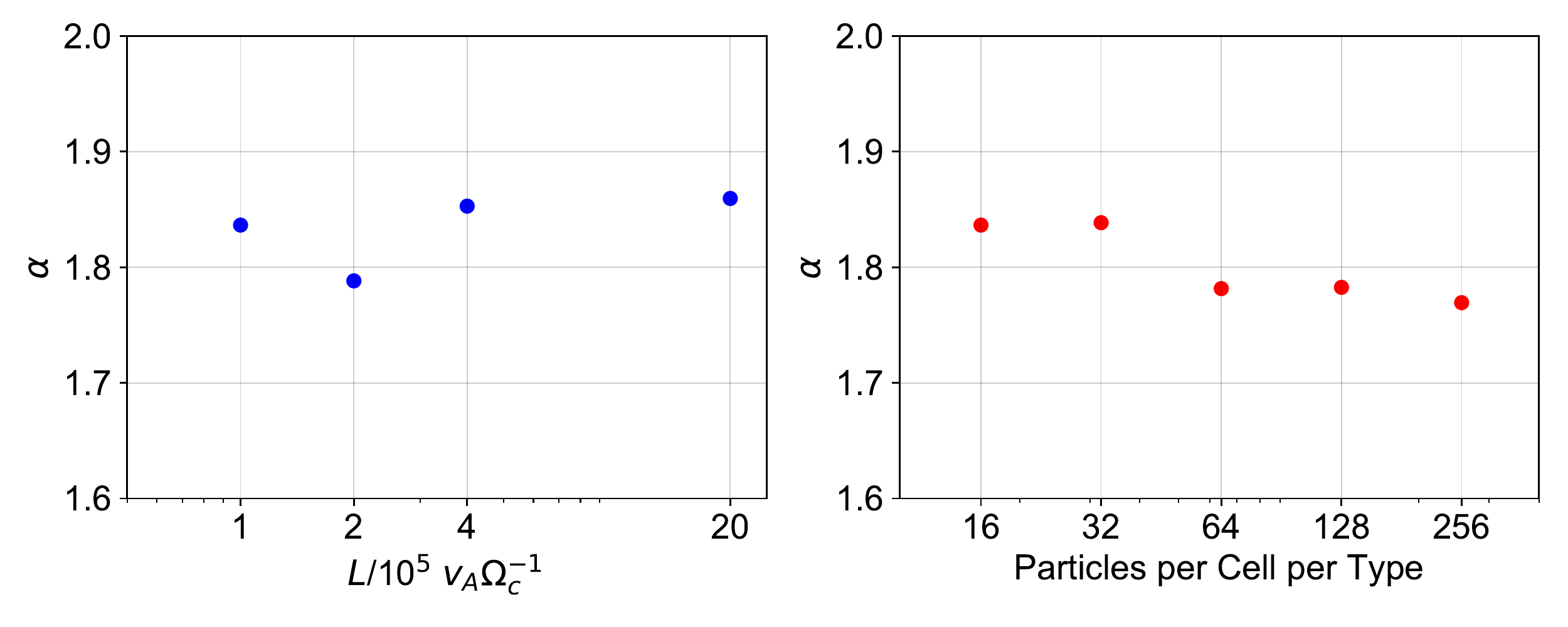,width=1.0\textwidth}}
\caption{Left: Saturation amplitude of \alf waves (Equation~\ref{eq:alpha}) as a function of box size $L$. Right: Saturation amplitude as function of particle number per cell per type. The saturation amplitudes are as directly measured from the simulations, computed as an average over the entire box at time $t$ = 10$^5$ $\Omega_c^{-1}$.}
\label{fig:alpha}
\end{figure*}

For a box size $L$, we require a region of diffuse ISM with length $L_{\mathrm{ISM}} > \lambda_{\mathrm{mfp}}$, the mean free path of cosmic rays to scattering by \alf waves. Using the definition of mean free path from Equation~\ref{eq:mfp} and taking a value of $p$~=~$p_0$~=~300~$v_A$, we require a box length,
\begin{equation}
    L \gtrsim 7.6 \times 10^2
    \left( \frac{\delta B_{\bot} (x)}{B_0} \right)^{-2} \: v_A \Omega_c^{-1}.
\end{equation}
\citetalias{Bai2019} used quasilinear theory to estimate the saturation amplitude of \alf waves (see \citetalias{Bai2019} Equation 25), finding that this amplitude lies in a range
\begin{equation}
    \underbrace{\frac{n_{\mathrm{CR,0}}}{n_i} \frac{\Delta v}{v_A}}_{\clap{non-relativistic}} \leq \left( \frac{\delta B_{\bot}}{B_0} \right)^2 \leq \underbrace{\frac{4}{3} 
    \frac{\langle p \rangle}{mc} \frac{n_{\mathrm{CR,0}}}{n_i} 
    \frac{\Delta v}{v_A}}_{\clap{ultra-relativistic}},
\end{equation}
where $\Delta v = v_D - v_A$ is the initial drift velocity and \nCRo/$n_i$ is the ratio of the equilibrium cosmic ray number density to the background ion number density. We can therefore parameterize the wave amplitude through a number $\alpha$,
\begin{equation} \label{eq:alpha}
    \alpha = \left( \frac{\delta B}{B_0} \right)^2 \left(\frac{n_{\mathrm{CR,0}}}{n_i} \frac{v_D-v_A}{v_A} \right)^{-1}.
\end{equation}
The value of $\alpha$ is computed in Figure~\ref{fig:alpha} by choosing the fiducial parameters of $v_D/v_A$ = 10, $n_{\mathrm{CR,0}}/n_i$ = 10$^{-4}$ and no damping throughout the box. This parameter is insensitive to box size and number of particles per cell. Taking the average value, $\alpha$ = 1.83, the minimum box size is
\begin{equation}
    L \gtrsim 4.2 \times 10^2 \left( \frac{n_{\mathrm{CR,0}}}{n_i} 
     \frac{\Delta v}{v_A}
     \right)^{-1} v_A \Omega_c^{-1}.
\end{equation}
For the fiducial parameters, $L \gtrsim$4.6$\times$10$^5$~$v_A \Omega_c^{-1}$. Thus, our chosen box size of $L$~=~2$\times$10$^6$~$v_A \Omega_c^{-1}$ contains $\approx$4 mean free paths, which we find sufficient to ensure that CR transport is diffusive after multiple traversals of the ISM.

Just as the \alf wave saturation amplitude is independent of particle number, we have verified that the linear gyroresonant CRSI growth rate, wave power spectra, and quasilinear evolution of the distribution function for our fiducial number of 16 particles per cell per type are in agreement with shorter, much higher resolution runs with 256 particles per cell per type. Thus, even though the decrease in particle number adds Poisson noise, the physics of the instability is properly captured in our simulations for the chosen parameters. 

\end{document}